\documentclass{siamart190516}

\graphicspath{{./figures/}}

\usepackage{amsfonts}
\usepackage{graphicx}
\usepackage{epstopdf}
\usepackage{algorithmic}
\usepackage{amsfonts,amsmath,amssymb}
\usepackage{mathrsfs,mathtools}
\usepackage{enumerate}
\usepackage{graphicx,color}
\usepackage{xcolor}
\usepackage{hyperref}
\usepackage{esint}
\usepackage{bm}
\usepackage{commath}
\usepackage{esint}
\usepackage{titlesec}
\DeclareMathAlphabet{\mathpzc}{OT1}{pzc}{m}{it}

\usepackage{placeins} 
\usepackage{flafter} 

\usepackage{tikz}

\floatname{algorithm}{Procedure}

\usepackage[textsize=small]{todonotes}
\numberwithin{equation}{section}

\makeatletter
\def\eqnarray{\stepcounter{equation}\let\@currentlabel=\theequation
\global\@eqnswtrue
\tabskip\@centering\let\\=\@eqncr
$$\halign to \displaywidth\bgroup\hfil\global\@eqcnt\z@
  $\displaystyle\tabskip\z@{##}$&\global\@eqcnt\@ne
  \hfil$\displaystyle{{}##{}}$\hfil
  &\global\@eqcnt\tw@ $\displaystyle{##}$\hfil
  \tabskip\@centering&\llap{##}\tabskip\z@\cr}

\def\endeqnarray{\@@eqncr\egroup
      \global\advance\c@equation\m@ne$$\global\@ignoretrue}

\newtheorem{remark}[theorem]{Remark}
\numberwithin{equation}{section}

\def\RR{{\mathbb{R}}}

\def\Om{\Omega}

\def\pOm{\partial\Omega}

\def\bbR{{\mathbb{R}}}

\newcommand{\WDNote}[1]           
{\textcolor{blue}{#1}\marginpar{\textcolor{blue}{WD $\longleftarrow$}}}

\ifpdf
  \DeclareGraphicsExtensions{.eps,.pdf,.png,.jpg}
\else
  \DeclareGraphicsExtensions{.eps}
\fi

\newsiamremark{hypothesis}{Hypothesis}
\crefname{hypothesis}{Hypothesis}{Hypotheses}
\newsiamthm{claim}{Claim}

\headers{Bilevel Optimization, Deep 
Learning and Fractional Laplacian}{Harbir Antil, Zichao (Wendy) Di, and Ratna Khatri}

\title{Bilevel Optimization, Deep 
Learning and Fractional Laplacian Regularization with Applications in Tomography\thanks{Submitted to the editors DATE.
\funding{The first and third authors are partially supported by NSF grants DMS-1818772, 
DMS-1913004 and the Air Force Office of Scientific Research under Award NO: FA9550-19-1-0036. The third author is also partially supported by a Provost award at George Mason University under the Industrial Immersion Program.
The second author is partially supported by DOE Office of Science under Contract No.\ DE-AC02-06CH11357.}}}

\author{Harbir Antil\thanks{Department of Mathematical Sciences, George Mason University, Fairfax, VA 22030, USA. 
  (\email{hantil@gmu.edu},  \email{rkhatri3@gmu.edu}).}
\and Zichao (Wendy) Di\thanks{Mathematics and Computer Science Division, Argonne National Laboratory, IL, USA.
  (\email{wendydi@mcs.anl.gov}).}
\and Ratna Khatri\footnotemark[2]}

\usepackage{amsopn}

\ifpdf
\fi

\begin{document}

\maketitle

\begin{abstract}
In this work we consider a generalized bilevel optimization framework for solving inverse problems. We introduce fractional Laplacian as a regularizer to improve the reconstruction quality, and compare it with the total variation regularization. We emphasize that the key advantage of using fractional Laplacian as a regularizer is that it leads to a linear operator, as opposed to the total variation regularization which results in a nonlinear degenerate operator. Inspired by residual neural networks, to learn the optimal strength of regularization and the exponent of fractional Laplacian, we develop a dedicated bilevel optimization neural network with a variable depth for a general regularized inverse problem. We also draw some parallels between an activation function in a neural network and regularization. We illustrate how to incorporate various regularizer choices into our proposed network. As an example, we consider tomographic reconstruction as a model problem and show an improvement in reconstruction quality, especially for limited data, via fractional Laplacian regularization. We successfully learn the regularization strength and the fractional exponent via our proposed bilevel optimization neural network. We observe that the fractional Laplacian regularization outperforms total variation regularization. This is specially encouraging, and important, in the case of limited and noisy data. 
\end{abstract}

\begin{keywords}
  bilevel optimization neural network, fractional Laplacian regularization, deep residual learning, imaging science, tomographic reconstruction, inverse problems.
\end{keywords}

\begin{AMS}
65D18, 68U10, 62H35, 94A08, 35R11, 34K37, 65K10.
\end{AMS}

\section{Introduction.}
\label{s:intro}

Inverse problems 
appear in numerous scientific domains, such as medicine, geophysics, astronomy, computer vision, and imaging etc.. However, they are typically ill-posed, due to the limited data and imperfection of experiments, and require some form of regularization \cite{hamalainen2013sparse,girard1987optimal,lassas2004can,niinimaki2016multiresolution, hsieh2013recent}. 
Two key challenges are associated with solving a regularized inverse problem. The first is the choice of regularization. 
Among the most popular choices, the total variation regularization \cite{TV_Rudin,shen2002mathematical} is of edge-preserving nature. However, its non-differentiability  makes its usage numerically challenging. Another choice is the Tikhonov regularization \cite{Tikhonov1977}, which has a smoothing property. Each choice, however, comes with its 
own challenges such as nonlinearity, non-smoothness, over-smoothing etc.. The second associated challenge is to choose the strength of the regularization, usually dictated by the parameter $\mu$, for which there is no consensus. 

Recently, deep learning approaches such as Convolution Neural Networks (CNN) and Residual Neural Networks (RNN) have shown remarkable potential in image
classification and reconstruction where, often, the goal is to learn the whole regularizer \cite{Pock2017_MRI,Yang_DeepADMM_2016,Zhang2017}. These 
approaches, however, may not be robust in general \cite{McCann2017, Lucas2018}. Firstly, learning problems are usually nonconvex, and the local minima may be sensitive to the initialization of parameters and the choice of optimization method. Secondly, these approaches often do not incorporate the domain-specific knowledge of the system (e.g., the known solution features) directly into the network, for instance.  In addition, they often lack a mathematical justification. 
The main contributions of this paper are two-folds:
    \begin{itemize}
        \item[(a)] Extend the fractional Laplacian introduced in \cite{antilsoren} as a regularizer to 
              the general setting of a linear inverse problem. 
              
        \item[(b)] Instead of learning the entire regularizer, we consider a  bilevel optimization scheme to learn the strength of the regularization and the fractional exponent based on the prior knowledge of the system. More specifically, we set up a \textit{bilevel optimization neural network (BONNet)}. In this network, the upper level objective measures an expectation of the reconstruction error over the training data 
        while the lower level problem measures the regularized data misfit.  
    \end{itemize}
        
There are several existing attempts to take advantage of machine learning to improve the solution quality. The most common way is to explore neural network as a post-processing step to refine the solution obtained by base-line methods (e.g., iterative method or filtered back projection \cite{kak2002principles}), see also \cite{CNN_tomo_2017,shan2019CT}. 

Our approach is closely related to the methodology introduced in \cite{Pock2017_MRI}, in fact ours can be thought as a special case in the  case of total variation, where the authors consider a variational model for reconstruction of MRI data. The authors focus on a generalized total variation model (Fields of Experts model) and  also learn the underlying parameters. We emphasize that the main novelty in our paper is the use of fractional Laplacian \cite{Caf3,PRStinga_JLTorrea_2010a,antilsoren} as a regularizer and learning the fractional exponent with an  application to tomographic reconstruction. The fractional Laplacian introduces nonlocality and tunable regularity. 
Another type of parameter search strategy has been proposed in \cite{JChung_MIEspanol_2017a} where the authors consider Tikhonov-based regularizations, and propose a machine learning based strategy to learn the strength of regularization. Their scheme is based on the generalized singular value decomposition (GSVD), or its approximation, of the forward operator and the regularization operator pair. 
However, computing GSVD can be computationally challenging  \cite{Hansen_GSVD}. Our approach differs from the existing works as we propose to use the fractional Laplacian   as a regularizer, which is cheaper to evaluate, and allows us to enforce the prior knowledge of the sample features, including smoothness and sparsity. The fractional Laplacian has been successfully applied in image denoising \cite{antilsoren,HAntil_CNRautenberg_2018b}, geophysics \cite{CWeiss_BvBWaanders_HAntil_2018a}, diffusion maps \cite{antil2018fractional}, biology \cite{AOBueno_DKay_VGrau_BRodriquez_KBurrage_2014a}, novel exterior optimal control \cite{HAntil_RKhatri_MWarma_2018a} etc.. We also emphasize that our proposed framework is flexible for it can easily
incorporate inequality constraints (on the optimization variables), which can be solved by a 
large number of existing solvers, and directly generalizes to other types of regularizations such as the $p$-Laplacian \cite{Bougleux2009,Cheng2019}. Therefore, our proposed framework brings machine learning closer to 
the traditional optimization. Notice that the machine learning algorithms are still in their infancy when it comes to handling constraints, see for instance 
\cite{JMagiera_DRay_JSHesthaven_CRohde_2019a} and the references therein.

The rest of the paper is organized as follows. In \cref{Sec:RegInvProb}, we introduce the mathematical formulation of the standard linear inverse problem with regularizers. In particular, we consider the
fractional Laplacian as a regularizer for inverse problems. We show a comparison of fractional Laplacian and total variation as regularizers for a tomographic reconstruction problem. 
\Cref{Sec:ParamSearch} is devoted to our proposed framework, i.e., the \textit{Bilevel Optimization Neural Network} to learn the optimal regularization strength, as well as the order of the fractional Laplacian. 
In \cref{numerics_tomo}, we provide further numerical experiments illustrating the application of BONNet to 
the tomographic reconstruction problem.

\section{Regularization in Inverse Problems. \label{Sec:RegInvProb}}
 The regression model for data misfit in inverse problems is given by
\begin{equation}\label{eq:a}
\min_{u} J(u) := \frac12 \|Ku-f\|_{L^2(\Omega)}^2,
\end{equation}
where $f: \Omega \mapsto \RR$ is a given function and $\Omega \subset \RR^n$ with $n\geq 1$ is a bounded domain. Here $K$ is the forward map, which we assume is a bounded linear operator on $L^2(\Omega)$ 
where the latter denotes the square integrable functions. Moreover, $u$ is the sample feature that we want to recover, or reconstruct. 
 The ill-posed nature of \cref{eq:a} makes it almost necessary to consider regularization in the wake of often noise-filled data; owing to the imperfections in the data gathering process. Therefore, we consider a regularized regression model to improve the solution quality. 
In a more general sense, let $\Om \subset \RR^n$ with $n \ge 1$ be a bounded Lipschitz domain with boundary 
$\pOm$, $f : \Omega \rightarrow \mathbb{R}$ be an $L^2(\Omega)$ function (given datum), 
$K : L^2(\Om) \rightarrow L^2(\Om)$ be a bounded linear operator, and $X$ be a Banach space. 
Then a standard regularized variational model is given by
\begin{equation}
\min_{u\in X_{ad} \subseteq X} J(u) := \frac12 \|Ku-f\|_{L^2(\Omega)}^2 + \mathcal{R}(u,\mu) ,
\label{eq:minp} 
\end{equation}
where $X_{ad}$ is a closed, convex, nonempty admissible set which is contained in the solution space $X$, and $u$ is the solution that we want to reconstruct or recover. Some examples of the operator $K$ for inverse problems in imaging science are the identity operator (image denoising problem) \cite{TV_Rudin}, convolution operator (image deblurring problem) \cite{MHintermueller_CNRautenberg_TWu_ALanger_2017a,MHintermueller_CNRautenberg_2017a}, and the Fourier or wavelet transforms \cite{JLStarck_ECandes_DLDonoho_2002a}. 
Therefore, in \cref{eq:minp}, the first term prevents the forward simulation from departing ``too far" away from $f$, thus it helps maintain the fidelity to $f$. In 
the absence of the second term ($\mathcal{R}(u,\mu)$), \cref{eq:minp} 
may be ill-posed \cite{Hansen1994}. The regularizer $\mathcal{R}(u,\mu)$ incorporates prior knowledge of the sample (like smoothness, sparsity, etc.), where $\mu$ balances the data misfit and the penalty enforced by the regularizer. Various choices of $\mathcal{R}(u,\mu)$ have been proposed in the literature.  In this work, we focus on the tomographic reconstruction problem, regularized with the fractional Laplacian,
and compare it against the total variation regularization. 

\subsection{Total Variation Regularization.} 

The penalty term for total variation (TV) regularization is given by 
\begin{equation}
\label{eq:TV}
\mathcal{R}(u,\mu) = \lambda \,\mbox{TV}(u) ,
\end{equation}
where $\mu = \lambda$ is a scalar. $\mbox{TV}(u)$ denotes the total variation semi-norm on $\Omega$   
and $X = BV(\Omega) \cap L^2(\Omega)$ where $BV(\Omega)$ denotes the set of functions of bounded variations \cite{AFP}. Formally speaking, TV$(u) := \int_\Omega |\nabla u|$ and as a result the corresponding Euler-Lagrange equations for \cref{eq:minp} are: Find $u \in X_{ad}\subset X$ such that 
\begin{equation}\label{eq:degeqn}
\left\langle -\lambda\; \mbox{div}\left(\frac{\nabla u}{|\nabla u|}\right) + K^*(Ku-f), \hat{u}-u \right \rangle_{X',X} \geq 0, \hspace{0.5cm}  \forall\:\: \hat{u}\in X_{ad}
\end{equation}
i.e., a nonlinear and possibly degenerate (due to $\displaystyle 1/|\nabla u|$) variational equation which 
is challenging to solve. We remark that $X'$ is the dual of $X$ and $K^*$ is the adjoint of $K$. Designing solvers for 
\cref{eq:degeqn} is still an active area of research \cite{bartels2017alternating}. 
The success of TV($u$) can be attributed to the fact that it prefers to fit shorter curves over the longer ones, thus avoids fitting noise and enforces sparsity. Additionally, it enforces much weaker regularity than the
$H^1$-regularization, i.e., when $\mathcal{R}(u,\mu) = \frac{\lambda}{2} \int_\Omega |\nabla u|^2$, with $\mu=\lambda$, and as a result it is possible to capture desirable sharp transitions in the reconstruction \cite{TV_Rudin}.
\subsection{Fractional Laplacian Regularization.}
The fractional Laplacian as a regularization for \cref{eq:minp} is given by,
\begin{equation}\label{eq:fraclap}
\mathcal{R}(u,\mu) = \frac{1}{2}\|\sqrt{\lambda}(-\Delta)^{\frac{s}{2}} u\|_{L^2(\Omega)}^2,
\end{equation}
where $\mu = (\lambda,s)$ is a vector. Moreover, with $0 < s < 1$, and $(-\Delta)^s$ denoting the fractional power of the classical Laplacian defined for instance in a spectral sense \cite{PRStinga_JLTorrea_2010a,antilsoren}. We remark that such a regularization enforces a reduced smoothness than $H^1$-regularization. The extent of the smoothness is dictated by the fractional power $`s$'. 
The key advantage of using this regularization is that the resulting Euler-Lagrange equation  for \cref{eq:fraclap} are: Find $u \in X_{ad}$ 
\begin{equation}\label{eq:degeqn_f}
\left\langle \lambda\; (-\Delta)^s u + K^*(Ku-f),\hat{u}-u \right\rangle \geq 0, \quad\quad \forall\, \hat{u} \in X_{ad}
\end{equation}
i.e., a variational equation with a linear operator. Such a problem has a unique 
solution in the fractional order Sobolev space $X = H^s(\Omega)$ \cite{kinderlehrer1980}. 
This regularization has been applied 
successfully in image denoising \cite{antilsoren}
(with $K = I$, but with $u \in X$, instead of $X_{ad}$, as a result \cref{eq:degeqn_f} becomes an equality). 
The success of this regularizer can be attributed to the fact that
when $s < \frac12$, the fractional Sobolev-space $H^s(\Omega)$ is larger than
$BV(\Omega) \cap L^\infty(\Om)$, see \cite{antilsoren}. As a result, we are solving a 
minimization problem over a larger space 
to achieve ``better'' results.

\subsection{Tomographic Reconstruction.\label{tomo}}

\begin{figure}[h!]
\includegraphics[width = 0.27\textwidth]{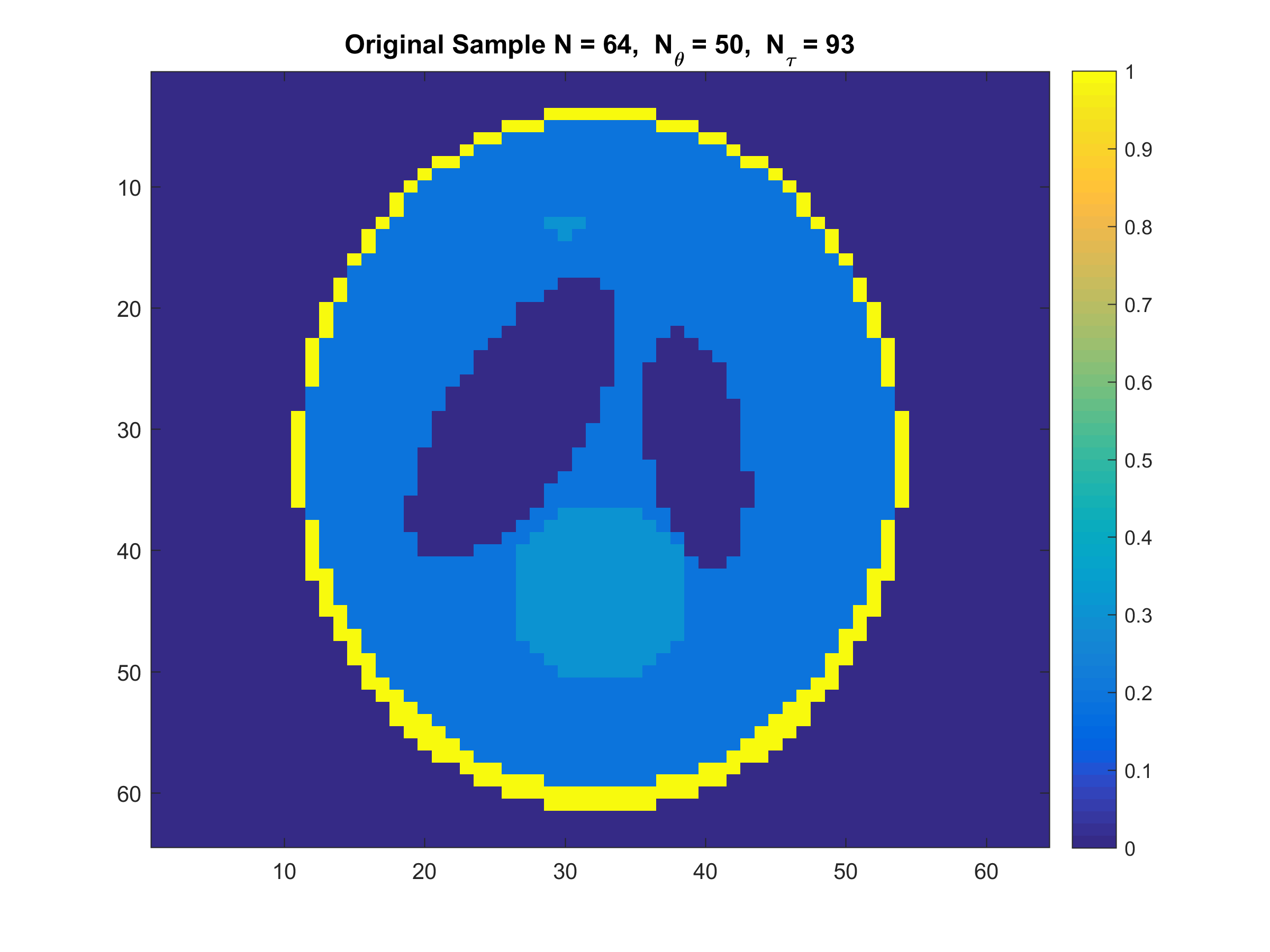} 
\includegraphics[width = 0.4\textwidth]{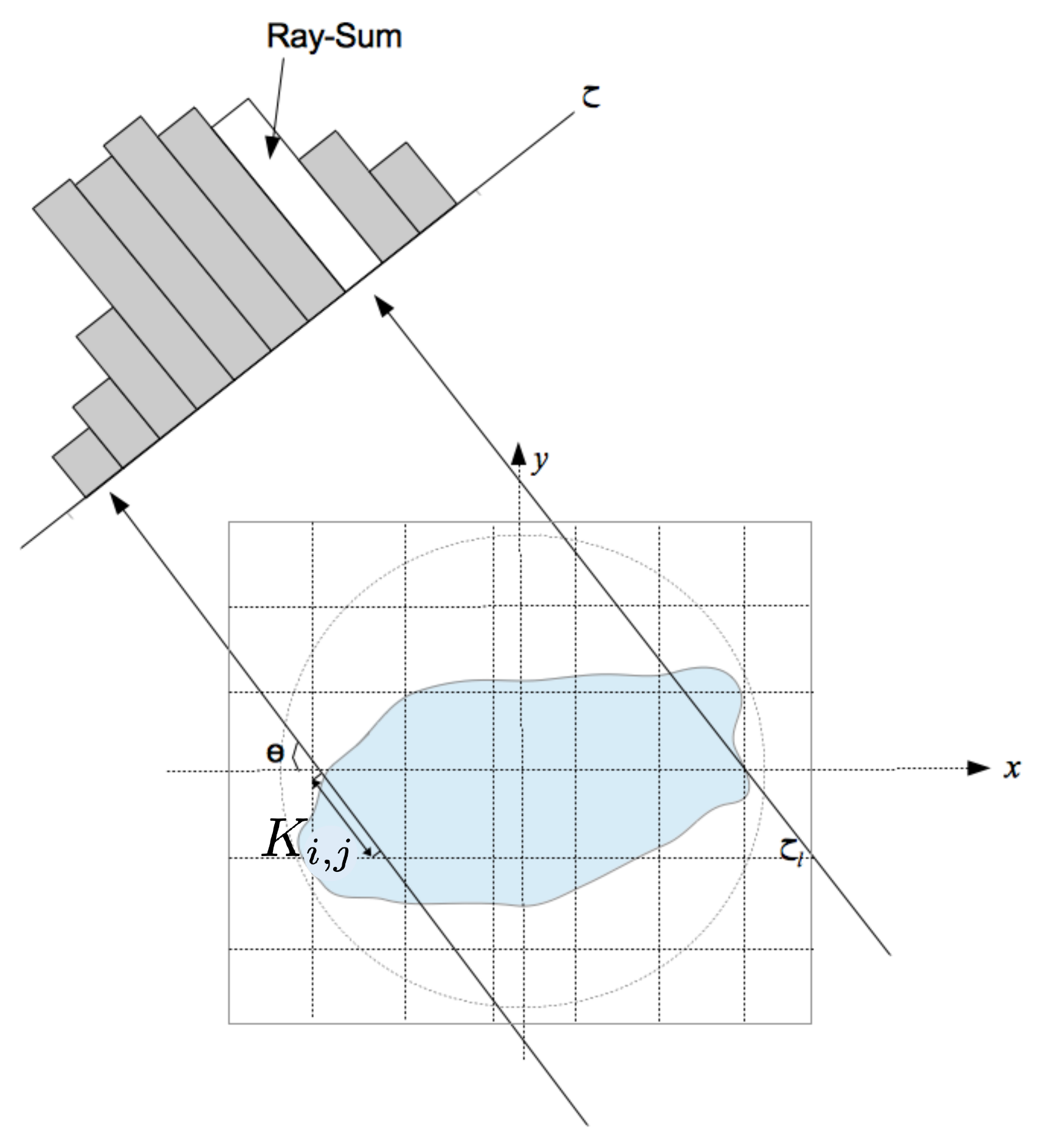}
\includegraphics[width = 0.23\textwidth, height = 0.27\textwidth]{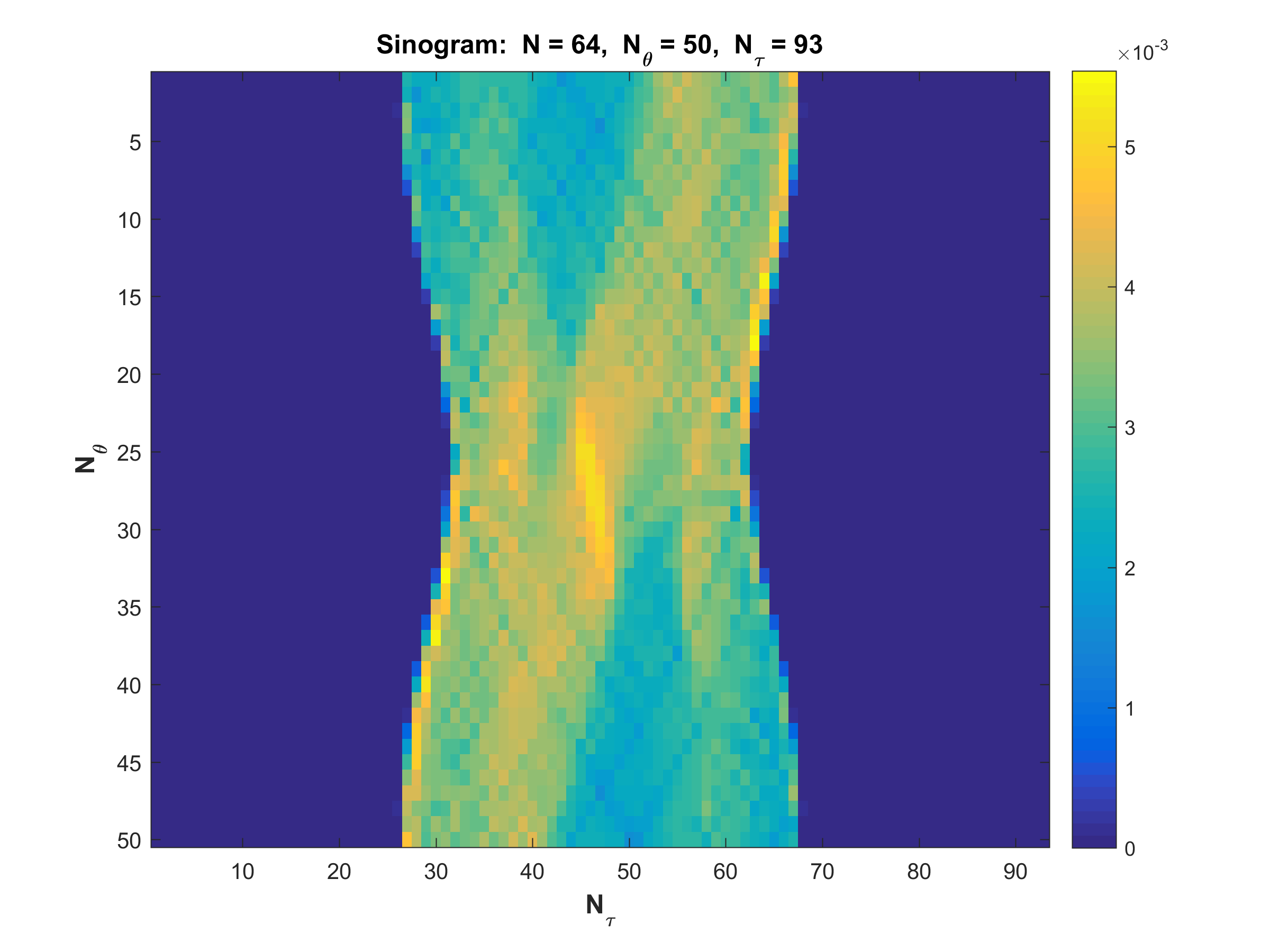}
\caption{
Geometric sketch of X-ray tomography (\textit{middle}) which maps the sample (\textit{left}) from the $(x,y)$ space to the sinogram (\textit{right}) on the $(\tau,\theta)$ space.}
\label{tomo_process}
\end{figure}

Tomographic reconstruction is a noninvasive imaging technique with the goal of recovering the internal characteristic
of a 3D object using a penetrating wave. It has shown revolutionary impact on various fields including physics, chemistry, biology, and astronomy. In a tomographic scan, a beam of light (e.g., X-ray) is projected onto the object to generate a 2D representation of the internal information along the beam path. By rotating the
object, a series of such 2D projections are collected from different angles
of view, collectively known as a sinogram (measurement data $f$), which can then be used to recover the internal characteristics (e.g., the attenuation coefficient) of the object \cite{wendyXray2016} (see \cref{tomo_process}).  
However, the limited data, due to the discrete nature of the physical experiment and dosage limits, makes the reconstruction problem ill-posed, i.e., many local minima exist for the objective function which is used to describe the discrepancy between the forward model and the measurement data. For illustration purpose, we confine ourselves to reconstruct 2D objects. 
The mathematical foundation of tomography is the Radon transform \cite{radon_determination_1986}, for which $K$ is defined as, 
\begin{equation}
\label{eqn:radon}
 Ku(\tau, \theta) := \int_{-\infty}^\infty \int_{-\infty}^\infty u(x, y) \delta(\tau - x\cos \theta - y \sin \theta) \: dx \: dy,
\end{equation}
where $u: \bbR^2 \mapsto \bbR$ is compactly supported on a bounded domain 
$\Omega \subset \mathbb{R}^2$ 
and $\delta$ is the Dirac mass, $\tau \in[0,\infty)$ and $\theta \in [0,2\pi)$ define the line of the beam path in a restricted domain. 
In practice, we can not recover the object at all points in space. Instead, we discretize $\Omega$ as $N \times N$ uniform pixels. Given $N_\theta$ number of angles and $N_\tau$ number of discrete beamlets,  our goal is to recover the piecewise constant approximation (on each pixel) $u\in \bbR^{N^2}$. Correspondingly, the discrete form of operator $K$ is the matrix ${\bf K} = (k_{i,j})_{i,j=1}^{N_\theta N_\tau,N^2}$ where the entries $k_{i,j}$ 
denote the contribution of $j$th pixel of $u$ to the $i$th component of the generated data. 

\subsection{Comparison of Fractional Laplacian with TV for Tomographic Reconstruction.}

To show the benefit of fractional Laplacian, we compare its performance against TV regularizer on a model problem. For now, we 
use common criterion to choose $\lambda$ and a fixed fractional exponent $s$ for this preliminary comparison. The rigorous computation of optimal $(\lambda,s)$ will be part of a forthcoming discussion.

\begin{figure}[h!]
\centering
\includegraphics[width = 0.49\textwidth , height=0.4\textwidth]{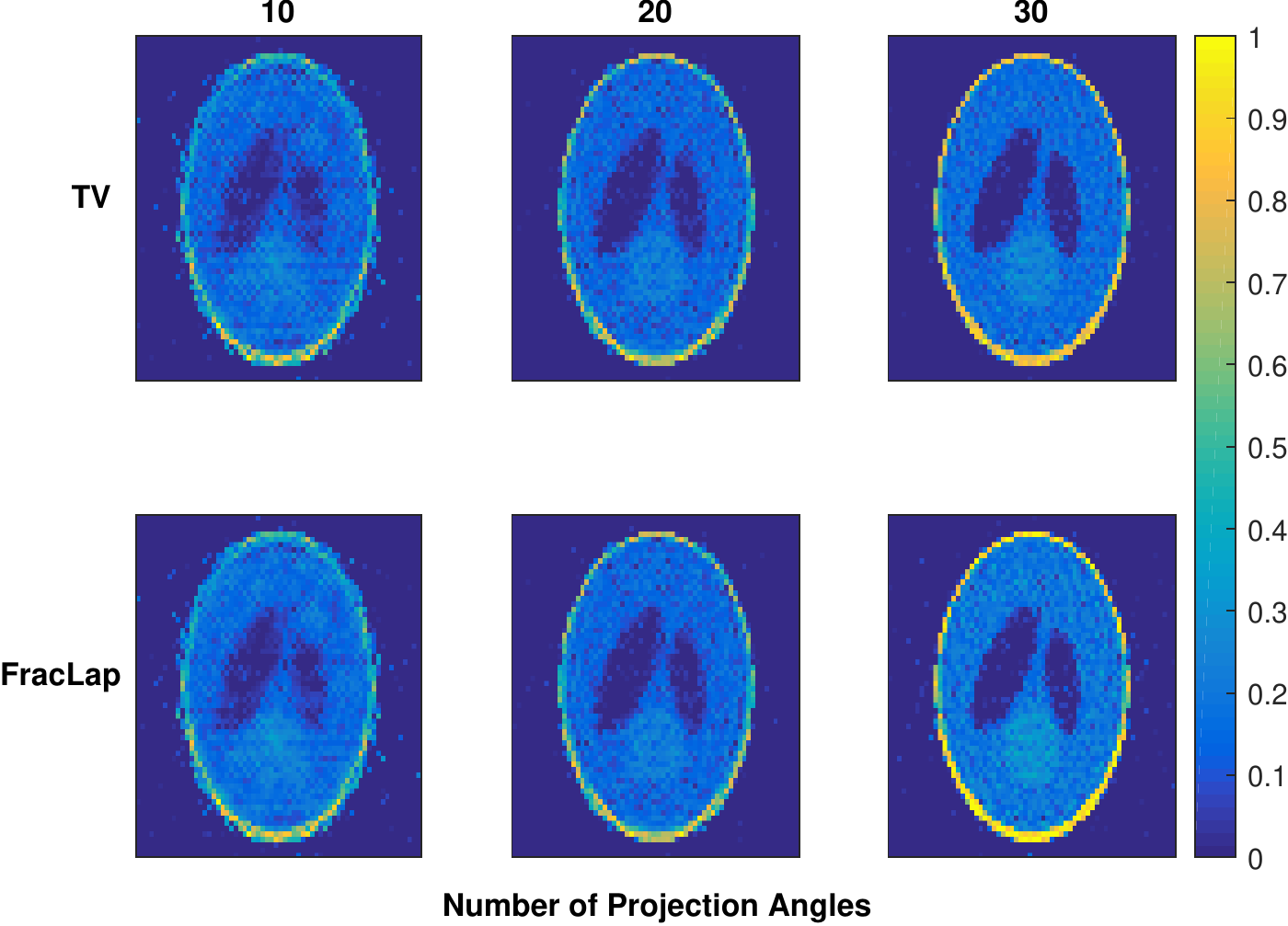}
\includegraphics[width = 0.49\textwidth , height=0.4\textwidth]{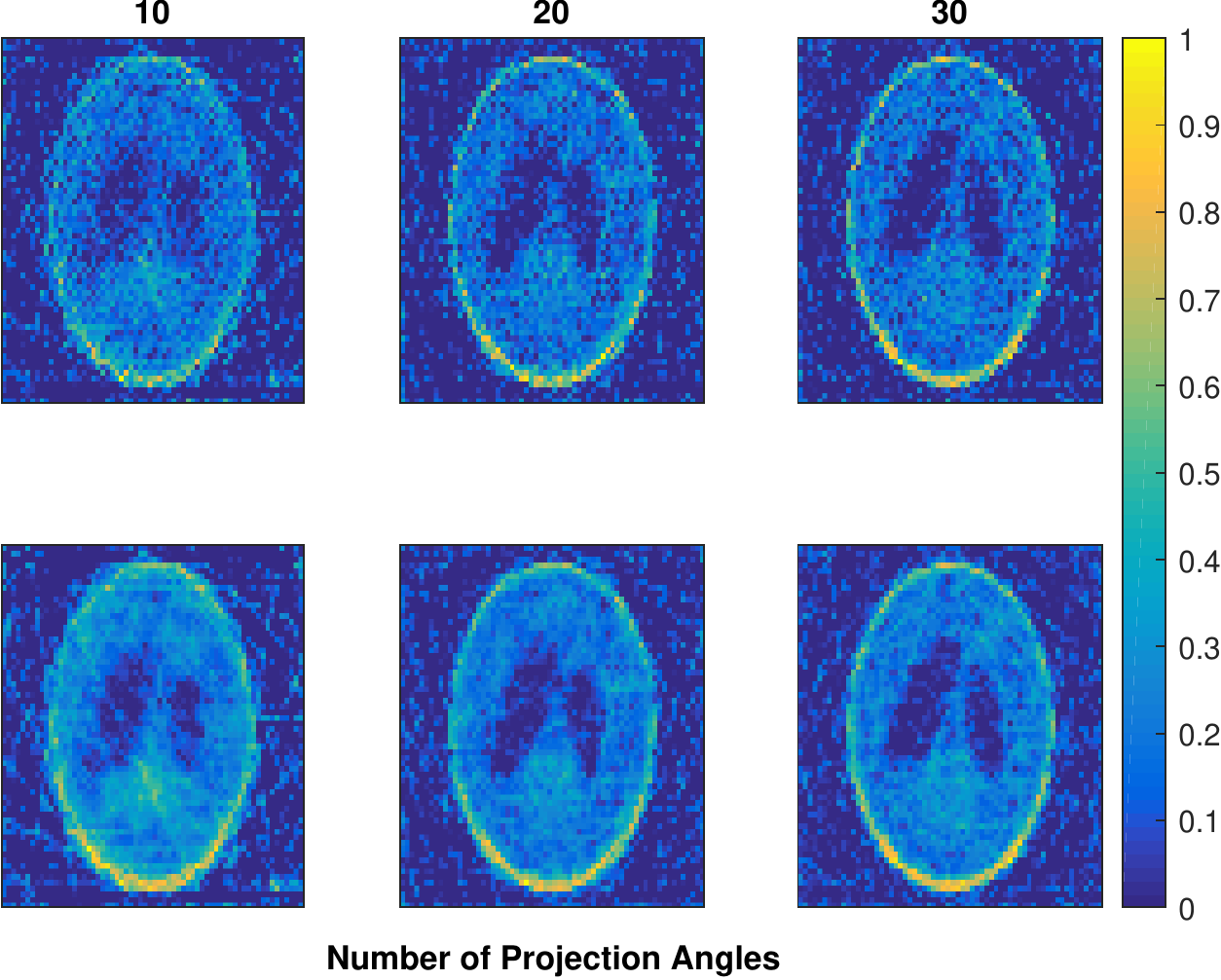}
\caption{ \small
Tomographic reconstructions based on the total variation regularization (\textit{row 1}) and fractional Laplacian (with $s=0.4$, \textit{row 2}) for data without noise (\textit{left}) and with $0.1\%$ noise (\textit{right}). The fractional Laplacian outperforms the total variation regularization in recovering finer features as well as in retaining high intensity regions, specially when the data is noisy and highly under-sampled.}
\label{fig:FracTVCompold}
\end{figure}

We choose our test problem as the tomographic reconstruction. First we synthetically generate the tomographic measurements of the sample $u$ by taking its discrete Radon transform, which gives us the data $f$. The sample $u$ and its corresponding sinogram $f$ are illustrated in \cref{tomo_process}. To get the noisy data, we add $0.1\%$ Gaussian noise to $f$. More details on tomographic reconstruction is provided in \cref{numerics_tomo}. Next we show the reconstructions based on the two regularizers, namely the fractional Laplacian \cref{eq:fraclap} and the total variation \cref{eq:TV} in \cref{fig:FracTVCompold}. 
The \textit{left} panel corresponds to  reconstructions based on sinogram $f$ without noise, and the \textit{right} panel corresponds to  reconstructions based on noisy $f$. \textit{Rows 1} and \textit{2} pertain to total variation and fractional Laplacian regularization, respectively. 

In the absence of noise, the reconstructions based on both regularizers are comparable. However, noiseless data does not depict a realistic situation 
\cite{Colton2013_InverseCrime}. 
In reality, the actual experimental data is always noisy due to the imperfections in the data acquisition 
process. We note that for noisy  data, particularly for the fewer projection case with $N_\theta = 10$ angles, fractional Laplacian regularization  gives better reconstructions than the total variation regularization. This can be specifically seen in \cref{fig:FracTVCompold} (\textit{right panel, row 2}) where finer features are better recovered 
e.g. the small circle at the bottom.  
 However, to fully explore the potential of regularization technique, the well-known challenge is to find the appropriate regularization strength $\lambda$ to optimally balance the trade-off between data misfit and prior knowledge enforcement.  
In the case of fractional Laplacian regularization, the exponent $`s$' only complicates the parameter choice further. 

For the reconstructions in \cref{fig:FracTVCompold}, given a wide range of values for $\lambda \in [1\times 10^{-18},10]$, we arbitrarily fix $s=0.4$, 
and solve the minimization problem \cref{eq:minp} using an inexact truncated-Newton method for bound-constrained problems \cite{Nash200045}. The optimal value of $\lambda$ is then chosen using a combination of L-curve criterion \cite{LcurveHansen} and the lowest $\ell_2$-norm of the reconstruction error compared to the ground truth. When L-curve criterion fails, we solely rely on the lowest $\ell_2$-norm. In our experience, this behavior is true for both TV and fractional Laplacian.  As a result, the optimal values of $\lambda$ for these tests is found to be in the range $[1\times 10^{-10},1]$. This procedure of finding an optimal $\lambda$ is labor-intensive, and 
requires access to the true solution, which is not available in practice. We remark that, to our experience, L-curve is efficient (not necessarily optimal) only in the case of strongly convex regularization which is definitely not the case with fractional Laplacian when $`s$' is also considered as a regularization parameter (non-convex with respect to $`s$'). L-curve criterion requires many different trial values of $\lambda$, along with a good guess of the interval to locate the corner of the L-curve.  
This requires a lot of human-intervention and fine-tuning. Furthermore, the regularized solution obtained by the $\lambda$ predicted by L-curve sometimes fails to converge to the true solution \cite{Vogel_Lcurve}. 

The next section addresses the issue of finding the optimal regularization parameters by proposing a deep bilevel optimization neural network, where the choice of regularizer is flexible.

\section{Parameter Learning via Bilevel Optimization Neural Network \label{Sec:ParamSearch}.}

Parameter search lies at the core of optimization. In particular, we seek parameters corresponding to the strength of regularization, which is a persistent challenge in the scientific community. To this end, we introduce a learning based approach as adverted in \cref{s:intro}.  We first state a generic bilevel optimization problem, 
\begin{equation}\label{eq:bilevel}
\begin{aligned}
&\min_{\mu \in \mathcal{M}_{ad}} \phi(\mu) \\
\mbox{s.t.} \quad  \min_{u\in X_{ad}} J(u,\mu) &:= \frac12 \|Ku-f\|_{L^2(\Omega)}^2 + \mathcal{R}(u,\mu) ,
\end{aligned}  
\end{equation}
where $\mathcal{M}_{ad}$ is a closed convex and nonempty admissible set for $\mu$. 

In \cref{rnn_phi}, motivated by \cite{Pock2017_MRI}, we present a machine-learning based approach to learn the regularization strength for a generic choice of regularizer. One of the key novelty of this paper is to use fractional Laplacian as the regularizer.
Notice that the lower level problem  \cref{eq:minp} in \cref{eq:bilevel} can be solved using existing techniques.

\subsection{Bilevel Optimization Neural Network (BONNet). \label{rnn_phi}}
Recently, deep residual learning has received a tremendous amount attention in machine learning for its immense potential to overcome the challenges faced by the traditional deep learning architectures, such as training complexity and vanishing gradients. These are resolved by adding skip connections, which transfer information between the layers \cite{resnet_HE2016}. Deep residual learning has enabled remarkable progress in imaging science \cite{resnet_HE2016,wu2018deep_imaging,CNN_tomo_2017}, biomedical applications \cite{lee2018deepres_bio,chen2018voxresnet_bio,Pock2017_MRI}, satellite imagery, remote sensing \cite{Tai_2017_CVPR,zhang2018missing,bischke2017detection} etc.. In our work, we use the power of deep learning to learn the regularization parameter $\mu$ which, for instance, contains the strength $\lambda$ and the fractional exponent $`s$'. 
We propose a dedicated deep bilevel optimization neural network to learn the regularization parameters. Our goal is to solve \cref{eq:bilevel} for which we seek our modeling inspiration from \cite{Pock2017_MRI}, and define $\phi(\mu)$ as the average mean squared error over $m$ distinct samples, i.e.
\[
\phi(\mu):= \frac{1}{2m} \sum\limits_{i=1}^{m} \| u^{(i)}(\mu) - u_{true}^{(i)}\|_{L^2(\Omega)}^2,
\]
where $u(\mu)$ solves the lower level problem in \eqref{eq:bilevel}, and corresponds to the sample characteristic that we wish to recover or reconstruct. Moreover, $u_{true}$, as the name suggests, is the \textit{known} true solution. 

We emphasize a few novelties of this work: first, our proposed network works directly on the data space, as opposed to the image space as a post-processing step as in \cite{ shan2019CT,CNN_tomo_2017}. Second, it generalizes to any bounded linear operator $K$ (the forward map; which defines the physics of the underlying system) and any $\mathcal{R}(u,\mu)$ (the regularization function; which allows us to incorporate the domain-specific knowledge of the solution). Third, we propose the use of fractional Laplacian as a regularizer with tunable regularity/smoothness. 
We also show how to integrate this choice of regularization into the BONNet architecture. We remark that fractional Laplacian introduces nonlocality in BONNet, which is challenging from both analytical and computational point of view. 

We first define the notion of a generalized regularizer and the projection map that we will be using to define the BONNet architecture.
\begin{itemize}
\item \textbf{Generalized Regularizer.}
Let $u(\mu)$ be the solution of the inner problem in \cref{eq:bilevel} which depends on $\mu$. 
Let $T:= T(\mu,u(\mu))$  be the action of some linear or nonlinear operator acting on $u(\mu)$, and $\sigma:= \sigma(T)$ be the \textbf{activation function} which acts pointwise on its argument. Then, we define a \textbf{generalized regularizer} as,
\begin{equation}\label{eq: Rgen}
\mathcal{R}(\sigma(T)) = \frac{1}{2} \| \sigma(T(\mu,u(\mu)))\|^2_{L^2(\Omega)}.
\end{equation}
Notice that in the case of \cref{eq:TV}, $\mu = \lambda$,  $T(\mu,u(\mu)) = 2\lambda |\nabla u |$, and $\sigma(T) = \sqrt{T}$. In the case of \cref{eq:fraclap}, $\mu = (\lambda,s)$, $T(\mu,u(\mu)) = \sqrt{\lambda} (-\Delta)^\frac{s}{2} u$, and $\sigma(T) = T$. 
Then, for $m$ distinct samples, we can write our inner minimization problem \cref{eq:minp} with a generalized regularizer as an average over $m$ samples,
\begin{equation}\label{inner_avg_prob}
\min \limits_{u \in X_{ad}} \: J(u,\mu) := \frac{1}{2m}\sum_{i=1}^m \: \left[ \|Ku^{(i)}-f^{(i)}\|_{L^2(\Omega)}^2 \:+\: \|\sigma\|_{L^2(\Omega)}^2 \right]
,\hspace{0.3cm} \mu \in \mathcal{M}_{ad}.
\end{equation}
To solve this inverse problem, we will employ derivative based methods such as projected gradient descent. The directional derivative of $J$ in a direction $h$ in \cref{inner_avg_prob} w.r.t $u$ in its variational form is; for each sample, $i = 1,...,m,$
\begin{small}
\begin{equation}\label{gradf}
DJ(u^{(i)},\mu)[h] = \frac{1}{m}\left[\left(K^*(Ku^{(i)}-f^{(i)}),h\right)_{L^2(\Omega)} + \left((\partial_{u^{(i)}} T\right)^*(\partial_T \sigma) \sigma,h)_{L^2(\Omega)}\right].
\end{equation}
\end{small}
\begin{remark}[\textbf{Regularization vs. Activation Function}]
{\rm
We notice a strong connection between the regularization function and the activation function used in machine learning architectures which is governed by \cref{eq: Rgen}. For instance, once we decide upon a choice of regularizer, the activation function is dictated by that choice. On the other hand, if we choose an activation function first, then the regularization is dictated by that choice. Thus, due to these parallels, regularization in a model seems to be similar in spirit to an activation function. 
}
\end{remark}

\item \textbf{{Solver: Projected Gradient Descent Method}\label{solver:pgd}}.
The choices of $X_{ad}$ and $\mathcal{M}_{ad}$ are problem dependent, for example, for tomographic reconstruction model, we let $X_{ad}:=\{ u \in X \ | \ u\ge 0 \}$. Moreover, we set 
$\mathcal{M}_{ad} := \Lambda_{ad}$ for total variation and $\mathcal{M}_{ad} := \Lambda_{ad} \times S_{ad}$ where $\Lambda_{ad} := \{ \lambda \in \mathbb{R} \ | \ \lambda \ge \epsilon_1 > 0\}$ 
and $S_{ad} := \{ s \in \mathbb{R} \ | \ 0 < \epsilon_2 \le s \le 1 - \epsilon_2\}$ for the fractional Laplacian. See \cref{s:admi} for more details on this application. 
In order to satisfy these constraints, we use the \textit{projected gradient descent method with line search} \cite{CTKellyOptim} to solve our inner and outer minimization problems in \cref{eq:bilevel}.  
Then, the projected gradient descent scheme for solving  (\ref{inner_avg_prob}), for a fixed $\mu$, $n$ iterations (\textit{depth of the network}), $\alpha$ as the line search parameter (i.e.\textit{ the learning rate}), $u_0$ as the initial guess, for the network layers (optimization iteration) $j = 1,...,n$, is given by 
\begin{equation}\label{eq:uj}
u_{j}^{(i)} = P_{X_{ad}}\left(u_{j-1}^{(i)}- \alpha \nabla_{u_{j-1}^{(i)}} J(u_{j-1}^{(i)},\mu)\right).
\end{equation}
where $P_{X_{ad}}(\cdot)$ denotes the projection on the admissible set $X_{ad}$, see \cref{s:admi} for more details on the tomographic reconstruction application. Note that, \cref{eq:uj} is also known as the \textit{forward propagation}. We are using $\nabla$ to denote the gradient and $D$ to denote the directional derivative (cf.~\cref{gradf}). 
 Now substitute the gradient from \cref{gradf} in \eqref{eq:uj} to arrive at,
\begin{equation} \label{minu_prob}
u_{j}^{(i)}= P_{X_{ad}}\Big(u_{j-1}^{(i)} -\frac{\alpha}{m} \big[K^*(Ku_{j-1}^{(i)}-f^{(i)})+(\partial_{u_{j-1}^{(i)}} T)^*(\partial_T \sigma) \sigma\big]\Big). 
\end{equation}
To compute the learning rate $\alpha$, we use line search for projected gradient descent as described in \cite[pg. 91]{CTKellyOptim}.
\end{itemize}
Putting it all together, we now describe our proposed \textbf{BONNet} architecture. 
Suppose we have $m$ distinct samples, and $n$ layers in our network. Let $u_{true}^{(i)}$ and $f^{(i)}$ be the known true solution and its corresponding experimental data for the $ith$ sample, with $i=1,...,m$. Then, we formulate our bilevel supervised learning problem as; for $j = 1,...,n$,
\begin{equation}\label{minlam_prob}
\begin{aligned}
&\min \limits_{\mu \in \mathcal{M}_{ad}} \phi(\mu)= \frac{1}{2m} \sum\limits_{i=1}^{m} \| u_n^{(i)}(\mu) - u_{true}^{(i)}\|_{L^2(\Omega)}^2 \\
\text{s.t.} \: \: u_{j}^{(i)}= \: &P_{ X_{ad}}\left(u_{j-1}^{(i)} -\frac{\alpha}{m} [K^*(Ku_{j-1}^{(i)}-f^{(i)})+(\partial_{u_{j-1}^{(i)}} T)^*(\partial_T \sigma) \sigma]\right).
\end{aligned}
\end{equation}
To solve the outer level problem for $\mu \in \mathcal{M}_{ad}$ 
we again use the projected gradient descent method, as described
above, with learning rate $\beta$ and $q$ iterations, 
\begin{equation} \label{lam_solve}
\mu_{l+1} = P_{\mathcal{M}_{ad}}\left(\mu_l-\beta \,\nabla_{\mu_l} \phi(\mu_l)\right), \hspace{1cm} l=0,...,q-1 , 
\end{equation}
where $P_{\mathcal{M}_{ad}}(\cdot)$ is the projection onto the admissible set. It then remains to evaluate $\nabla_{\mu_l} \phi(\mu_l)$. After applying the chain rule, we obtain that
\begin{equation}\label{gradPhi}
\nabla_{\mu_l} \phi(\mu_l) =  
\frac{1}{m}\sum\limits_{i=1}^{m} \int_{\Omega} (u_n^{(i)} - u_{true}^{(i)})\left.\frac{du_n^{(i)}}{d\mu}\right|_{\mu=\mu_l} d\Omega.
\end{equation}
As noted earlier, the most challenging part of this network is the computation of sensitivity of $u$ w.r.t. $\mu$, because at each network layer, $u$ depends on the previous iterate, as well as $\mu$, as can be seen in the lower level problem in \cref{minlam_prob}. We evaluate  $\left. \frac{du_n^{(i)}}{d\mu}\right|_{\mu=\mu_l}$in \cref{gradPhi} by implicit differentiation. This 
results in an iterative system of equation that we need to solve. For each sample index $`i$', it is explicitly derived as follows, for $j = 1,...,n$
\begin{equation}\label{dulam}
\left.\frac{du_j}{d\mu}\right|_{\mu=\mu_l} = \left. \frac{\partial u_j}{\partial u_{j-1}}\cdot\frac{d u_{j-1}}{d\mu}\right|_{\mu=\mu_l}+ \left. \frac{\partial u_j}{\partial \mu}\cdot \frac{d \mu}{d\mu}\right|_{\mu=\mu_l},
\end{equation}
where,
\begin{equation}\label{parduu}
\begin{aligned}
\frac{\partial u_j}{\partial u_{j-1}} = I -\frac{\alpha}{m} &\left[K^*K +  \frac{\partial }{\partial u_{j-1}}\Big(\frac{\partial T}{\partial u_{j-1}}\Big)^* \Big(\frac{\partial \sigma}{\partial T}\Big)^* \sigma +\Big(\frac{\partial T}{\partial u_{j-1}}\Big)^* \frac{\partial }{\partial u_{j-1}}\Big(\frac{\partial \sigma}{\partial T}\Big)^* \sigma + \right. \\
&\Big(\left.\frac{\partial T}{\partial u_{j-1}}\Big)^* \Big(\frac{\partial \sigma}{\partial T}\Big)^*\Big(\frac{\partial \sigma}{\partial T} \cdot \frac{\partial T}{\partial u_{j-1}}\Big)\right],
\end{aligned}
\end{equation}
and,
\begin{equation} \label{pardulam}
\begin{aligned}
\frac{\partial u_j}{\partial \mu} = -\frac{\alpha}{m}&\left[\Big(\frac{\partial }{\partial \mu} \Big( \frac{\partial T}{\partial u_{j-1}} \Big)^*\Big) \Big(\frac{\partial \sigma }{\partial T}\Big)^* \sigma  + \Big(\frac{\partial T}{\partial u_{j-1}}\Big)^* \Big(\frac{\partial }{\partial \mu } \Big( \frac{\partial \sigma}{\partial T} \Big)^*\Big) \cdot \sigma + \right.\\
&\left. \Big( \frac{\partial T}{\partial u_{j-1}} \Big)^*\Big( \frac{\partial \sigma}{\partial T} \Big)^* \cdot \frac{\partial \sigma}{\partial T} \frac{\partial T}{\partial \mu}\right].
\end{aligned}
\end{equation}
Substituting \cref{parduu} and \cref{pardulam} in \cref{dulam} yields the sensitivity of $u$ w.r.t. $\mu$. Now that we have the key architecture of the deep BONNet, we divide our network into a \textit{training} phase and a \textit{testing} phase, as is common in a standard machine learning framework. During the \textit{training} phase, we solve the bilevel optimization problem \cref{minlam_prob} to learn the regularization parameters, and during the \textit{testing} phase we only solve the inner problem  in \cref{minlam_prob} using the regularization parameters learned from the training phase. The training phase can be carried out offline (i.e. in advance), and testing phase can be carried out online (i.e. as the experimental data becomes available).

\subsubsection{General Framework of BONNet.}
 We summarize the training and testing phases of our deep  BONNet architecture as follows:
\begin{itemize}
\item \textbf{\textit{Training Phase} (\cref{train_phase_alg}).}
In this phase, we pass in $m$ training samples $\left\{u_{true}^{(i)},f^{(i)}\right\}_{i=1}^{m}$
to learn the \textit{optimal} $\mu$ which we denote by $\mu^*$. 
The depth of the deep BONNet at the \textit{training} phase is $`q$ sets of $n$ layers'. This phase can be carried out offline. 
\begin{algorithm}[h!]
\caption{Training Phase of BONNet}
\label{train_phase_alg}
\begin{algorithmic}[1]
\REQUIRE $\left\{u_{true}^{(i)},f^{(i)}\right\}_{i=1}^{m}$, $m$ training samples
\ENSURE $\mu^*$
\STATE Initialize $u_0$, $\frac{du_0}{d\mu}$  and $\mu_0$
\FOR {for $l = 0$ to $q-1$} 
\FOR{for $j = 1$ to $n$}
\STATE Compute $u^{(i)}$ and $\frac{du_n^{(i)}}{d\mu}$ for all $i = 1,...,m$:
\[
u_{j}^{(i)}= P_{X_{ad}} \Big(u_{j-1}^{(i)} -\frac{\alpha}{m} \big[K^*(Ku_{j-1}^{(i)}-f^{(i)})+(\partial_{u_{j-1}^{(i)}} T)^*(\partial_T \sigma) \sigma\big]\Big). 
\]
\COMMENT{Compute $\alpha$ using line search as discussed in \cref{solver:pgd}}
\[
\left.\frac{du_j^{(i)}}{d\mu}\right|_{\mu=\mu_l} = \left. \frac{\partial u_j^{(i)}}{\partial u_{j-1}^{(i)}}\cdot\frac{d u_{j-1}^{(i)}}{d\mu}\right|_{\mu=\mu_l}+ \left. \frac{\partial u_j^{(i)}}{\partial \mu}\cdot \frac{d \mu}{d\mu}\right|_{\mu=\mu_l}
\]
\COMMENT{See \cref{parduu,pardulam} for explicit expressions}
\ENDFOR
\STATE Compute the gradient of $\phi(\mu)$:
\[
\nabla_{\mu_l} \phi(\mu_l) =  
\frac{1}{m}\sum\limits_{i=1}^{m} \int_{\Omega} (u_n^{(i)} - u_{true}^{(i)})\left.\frac{du_n^{(i)}}{d\mu}\right|_{\mu=\mu_l} d\Omega,
\]
\STATE Update $\mu$:
\[
\mu_{l+1} = P_{\mathcal{M}_{ad}}\left(\mu_l-\beta \nabla_{\mu_l} \phi(\mu_l) \right).
\]
\COMMENT{Compute $\beta$ using line search as discussed in \cref{solver:pgd}}
\ENDFOR
\end{algorithmic}
\end{algorithm}

\item \textbf{\textit{Testing Phase} (\cref{test_phase_alg}).}
In this phase, we use the $\mu^*$ learned from the \textit{training} phase
and testing data $\left\{f^{(i)}_{test}\right\}_{i=1}^{m_{test}}$ 
to \cref{test_phase_alg}. 
The depth of the network at the \textit{testing} phase is $n_{test}$ layers. This phase can be carried out online, once the experimental data $f_{test}$ becomes available.  
\begin{algorithm}[h!]
\caption{Testing Phase of BONNet}
\label{test_phase_alg}
\begin{algorithmic}[1]
\REQUIRE $\mu^*, \left\{f^{(i)}_{test}\right\}_{i=1}^{m_{test}}$, $m_{test}$ testing samples
\ENSURE $u$
\STATE Initialize $u_0$
\FOR{for $j = 1$ to $n_{test}$}
\STATE Compute $u$ for all $i = 1,...,m_{test}$:
\[
u_{j}^{(i)}= P_{X_{ad}} \Big(u_{j-1}^{(i)} -\frac{\alpha}{m} \big[K^*(Ku_{j-1}^{(i)}-f^{(i)}_{test})+(\partial_{u_{j-1}^{(i)}} T)^*(\partial_T \sigma) \sigma\big]\Big). 
\]
\COMMENT{Compute $\alpha$ using line search as discussed in \cref{solver:pgd}}
\ENDFOR
\end{algorithmic}
\end{algorithm}
\end{itemize}

\begin{remark}[\textbf{Fixed vs. Variable Depth of BONNet.}]
\label{rem:fix}
{\rm
We remark that instead of specifying the number of layers when solving \cref{lam_solve} or \cref{minu_prob}, one could also specify a stopping criterion appropriate for the solver being used, which is our recommendation as well. This is more in the spirit of solving an optimization problem which converges to a solution. This implies that the layers of the deep BONNet, in this case, will be variable. In our numerical experiments, we have used the stopping criterion for projected gradient descent method as mentioned in \cite[pg. 91]{CTKellyOptim} for both $\mu$ and $u$. Also note that for \cref{minu_prob}, the number of layers in the testing phase ($n_{test}$) does not have to be equal to the number of layers in the training phase ($n$). In fact, $n<<n_{test}$ prevents the network from \textit{overfitting} of parameters to the training data. Furthermore, reconstruction at the testing phase can be progressively improved for structural fidelity, if needed, by using a larger $n_{test}$ (or a stricter stopping criterion). This allows for a trade-off between the quality of reconstruction and computational time. 
}
\end{remark}

\subsubsection{BONNet Framework for Fractional Laplacian and Total Variation Regularization.\label{Reg_examples}}

In the general framework of our proposed deep BONNet, for any bounded linear operator $K$, any choice of regularizer can be incorporated, as long as it is cast into the generalized regularizer framework \cref{eq: Rgen}. 
In \cref{Sec:RegInvProb}, we have proposed the use of fractional Laplacian as a regularizer, and have compared it with total variation regularization. We now show how to incorporate these regularizers into the deep BONNet, for a general $K$:

\begin{itemize}
\item[(a)] \label{fracLap_archit} \textbf{Fractional Laplacian Regularization.} Recall the fractional Laplacian regularization from \cref{eq:fraclap},
\[
\mathcal{R}(u,\mu) = \frac{1}{2}\|\sqrt{\lambda}(-\Delta)^{\frac{s}{2}} u\|_{L^2(\Omega)}^2,
\]
where $\mu = (\lambda,s)$ and $s \in (0,1)$. Then, to define the corresponding generalized regularizer \cref{eq: Rgen}, let $T(\mu,u(\mu)) := \sqrt{\lambda} (-\Delta)^\frac{s}{2} u$, and the activation function $\sigma(T) := T$. We omit the superscript $`i$' to improve readability. Then, after some simplifications, \cref{minlam_prob}, \cref{parduu}, and \cref{pardulam} become, for $j = 1,...,n$,
\[
u_{j}= P_{X_{ad}} \Big(u_{j-1} -\frac{\alpha}{m} \big[K^*(Ku_{j-1}-f)+ \lambda (-\Delta)^{s}u_{j-1}\big] \Big),
\]
\[
\frac{\partial u_j}{\partial u_{j-1}} =  I-\frac{\alpha}{m} K^*K-\frac{\alpha \lambda}{m} (-\Delta)^s,
\]
and
\begin{equation}\label{eq:us}
\frac{\partial u_j}{\partial \lambda} = - \frac{\alpha}{m} (-\Delta)^s u_{j-1} \quad \mbox{and} \quad 
\frac{\partial u_j}{\partial s} = -\frac{\alpha \lambda}{m}\frac{\partial}{\partial s}((-\Delta)^su_{j-1})
\end{equation}
which together give us the sensitivity of $u$ w.r.t. $\mu$ in  \cref{dulam}. Notice that the second equation in \eqref{eq:us} 
requires the sensitivity of fractional Laplacian $(-\Delta)^s$ with respect to $`s$'. This is a highly delicate object to handle. We shall reserve further details on this topic until the next section. 

\item[(b)] \textbf{Total Variation Regularization.} Recall the total variation regularization 
\[
\mathcal{R}(u,\mu) = \lambda \,\mbox{TV}_{\xi}(u) ,
\]
where $\mu = \lambda$, and we are using the  ``regularized'' total variation semi-norm,
\begin{equation}\label{eq:TVReg}
\text{TV}_{\xi}(u)=  \int_{\Omega} \sqrt{|\nabla u|^2_{\ell^2(\Omega)} + \xi^2}\: \partial \Omega .
\end{equation}
with $0<\xi \ll 1 $. We will omit the subscript $\xi$ from $\text{TV}_{\xi}$ for brevity. Then, to define the corresponding generalized regularizer \cref{eq: Rgen}, let $T(\mu,u(\mu)) := 2\lambda \text{TV}(u)$, and the activation function $\sigma(T) := \sqrt{T}$.  Then, after some simplifications, \cref{minlam_prob}, \cref{parduu}, and \cref{pardulam} become, for $j = 1,...,n$,
\[
u_{j}= P_{X_{ad}} \bigg(u_{j-1} -\frac{\alpha}{m} \Big[K^*(Ku_{j-1}-f) + \lambda \Big(-\text{div}\big(\frac{\nabla u_{j-1}}{\sqrt{|\nabla u_{j-1}|^2_{\ell^2(\Omega)} + \xi^2}}\big)\Big)\Big]\bigg),
\]
\begin{equation}\label{parduu_TV}
\frac{\partial u_j}{\partial u_{j-1}} =  I - \frac{\alpha}{m} K^*K + \frac{\alpha\lambda}{2m} \text{ div } \Big( \frac{\partial}{\partial u_{j-1}}\big(\frac{\nabla u_{j-1}}{\sqrt{|\nabla u_{j-1}|^2_{\ell^2(\Omega)} + \xi^2}}\big)\Big) ,
\end{equation}
and
\[
\frac{\partial u_j}{\partial \lambda} =-\frac{\alpha}{2m} \Big(-\text{div} \big( \frac{\nabla u_{j-1}}{\sqrt{|\nabla u_{j-1}|^2_{\ell^2(\Omega)} + \xi^2}}\big)\Big)^*,
\]
which together give us the sensitivity of $u$ w.r.t. $\mu$ in  \cref{dulam}. Again, we have omitted the superscript $`i$' to improve readability.
\end{itemize}

\section{Numerical Experiments of Tomographic Reconstruction. \label{numerics_tomo}}

In this section, we present several numerical experiments where we apply our proposed BONNet  
to a tomographic reconstruction problem. We have introduced tomographic reconstruction in 
\cref{tomo}. 
We demonstrate the results of BONNet with two regularizers, namely, the total variation and the proposed fractional Laplacian \cref{eq:fraclap}. 

All the computations are carried out using MATLAB R2015b on a Laptop with Intel Core i7-8550U Processor, with NVIDIA GeForce MX150 with 2 GB RAM. 
In view of \cref{rem:fix}, we run the proposed algorithm until a desired tolerance (tol) is met. At the testing phase we set tol $=1\times 10^{-5}$ and at the training phase we set tol $=1\times 10^{-3}$. Notice that the former is stricter than latter to avoid \emph{overfitting}.

For all the total variation experiments we set the regularization parameter $\xi$ in \cref{eq:TVReg} as $\xi = 1\times 10^{-5}$.  Moreover, the last term in \cref{parduu_TV} is simply approximated by discrete Laplacian.

The remainder of the section is organized as follows. First in \cref{s:prelim} we discuss the implementation details of fractional Laplacian and the admissible sets $X_{ad}$ and $\mathcal{M}_{ad}$. This is followed by two experiments in \cref{s:two_ex}.

\subsection{Preliminaries.}
\label{s:prelim}

Before we discuss the actual results, we state some  preliminary material. As mentioned in the paragraph following \cref{eqn:radon}, we discretize $\Omega$ as $N\times N$ uniform 
pixels. Then given $N_\theta$ number of angles and $N_\tau$ number of discrete beamlets, our goal is to recover 
$u \in \mathbb{R}^{N^2}$. We also recall that the discrete form of operator $K$ is the matrix 
${\bf K} = (k_{i,j})_{i,j=1}^{N_\theta N_\tau, N^2}$. All the integrals are computed using uniform quadrature and the differential operators are discretized using finite differences. We shall discuss the approximation of fractional Laplacian next.

\subsubsection{Numerical Approximation of Fractional Laplacian.}
\label{s:fracLapapprox}

In order to approximate the fractional Laplacian, we first discretize the Laplacian $(-\Delta)$ on a uniform stencil. We denote the resulting discrete matrix by $\mathbf{A}$. If the eigen-decomposition of $\mathbf{A}$ is
    \[
        \mathbf{A} = \mathbf{V} \mathbf{D} \mathbf{V}^{-1},
    \]
where $\mathbf{D} = (d_{i,j})_{i,j=1}^{N^2,N^2}$ with $d_{i,j} = 0$ if $i\neq j$, and $d_{i,i} = \zeta_i$ denotes the eigenvalues with columns of $\mathbf{V}$ containing the corresponding eigenvectors. Then the fractional power of $\mathbf{A}$ is given by,
    \[
        \mathbf{A}^s = \mathbf{V} \mathbf{G}(s) \mathbf{V}^{-1},
    \]
where $\mathbf{G}(s) = (g_{i,j}(s))_{i,j=1}^{N^2,N^2}$ is the diagonal matrix with $g_{i,j}(s) = 0$ if $i\neq j$ and $g_{i,i}(s) = \zeta_i^s$. From \cref{eq:us} we also recall that we need to approximate the variation of ${\bf A}^s$ with respect to $`s$'. A straightforward calculation gives 
    \[
        \frac{d}{ds} \mathbf{A}^s = \mathbf{V} \mathbf{H}(s) \mathbf{V}^{-1}
    \]
where $\mathbf{H}(s) = (h_{i,j}(s))_{i,j=1}^{N^2,N^2}$ is the diagonal matrix with $h_{i,j}(s) = 0$ if $i\neq j$ and $h_{i,i}(s) = \zeta_i^s \ln(\zeta_i)$.    

\subsubsection{Admissible Sets and Projection.}
\label{s:admi}

For tomographic reconstruction we let $X_{ad}:=\{ u \in X \ | \ u\ge 0 \}$. Moreover, we set $\mathcal{M}_{ad} := \Lambda_{ad}$ for total variation and $\mathcal{M}_{ad} := \Lambda_{ad} \times S_{ad}$ where $\Lambda_{ad} := \{ \lambda \in \mathbb{R} \ | \ \lambda \ge \epsilon_1 > 0\}$ and $S_{ad} := \{ s \in \mathbb{R} \ | \ 0 < \epsilon_2 \le s \le 1 - \epsilon_2\}$. We let $\epsilon_1 = \epsilon_2 = 10^{-15}$. 

\noindent Furthermore, the projection in \cref{minu_prob} onto the admissible set $X_{ad}$ is given by, for $z\in X$,
\begin{equation}  \label{ProjMap}
P_{X_{ad}}(z):= \text{max} \, \{ 0,z\} = \left\{
\begin{aligned}
&z \quad &\text{if}\quad z\ge 0,\\
&0 \quad &\text{if}\quad z< 0  . 
\end{aligned}
 \right.
\end{equation}
Formally, the ``derivative'' of this map is given by 
\[
\frac{d}{dt}\left( P_{X_{ad}}(z)\right):= \left\{
\begin{aligned}
&\frac{dz}{dt} \quad &\text{if}\quad z\ge 0,\\
&0 \quad &\:\text{if}\quad z< 0 . 
\end{aligned}
\right.
\]
For a rigorous definition of the generalized derivative of the \textit{max} function, see \cite{Clarke1975}. Similar projection
formulas are applicable for projection onto the set $\mathcal{M}_{ad}$.

\subsection{Experiments.}
\label{s:two_ex}

\begin{figure}[h!]
\begin{center}
\includegraphics[width = 0.35\textwidth , height=0.25
\textwidth]{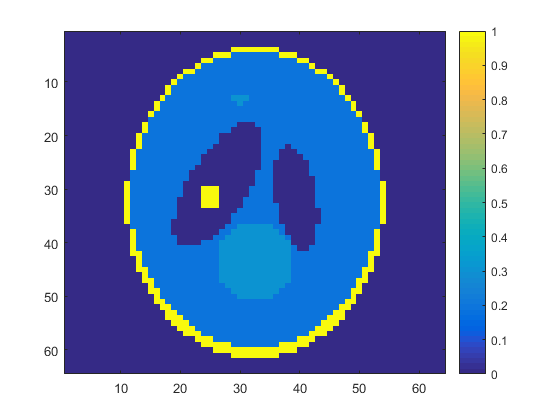}
\includegraphics[width = 0.35\textwidth , height=0.25
\textwidth]{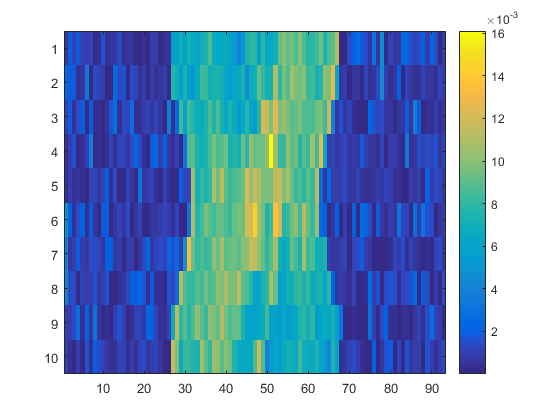}
\end{center}
\begin{center}
\includegraphics[width = 0.35\textwidth , height=0.25\textwidth]{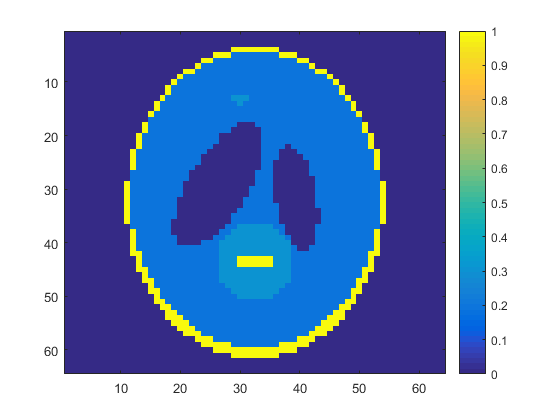}
\includegraphics[width = 0.35\textwidth , height=0.25
\textwidth]{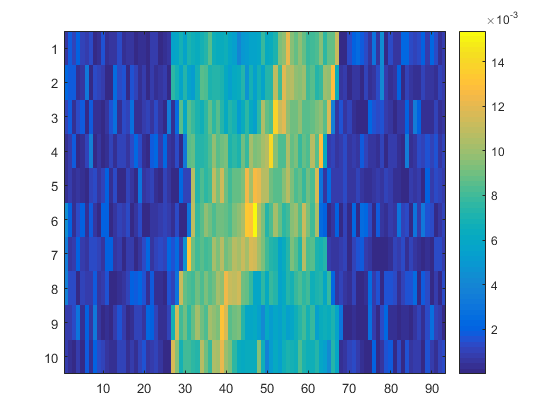}
\end{center}
\caption{Representative samples of Phantom ($u_{true}$) used (\textit{left}) to generate the synthetic data (noisy sinogram $f$) (\textit{right}) for training (\textit{Row 1}) and testing (\textit{Row 2}).\label{train_test_samp}}
\end{figure}

We begin by generating the synthetic data. We create $30$ distinct $64 \times 64$ samples (i.e. $N=64$), which are variations of the Shepp-Logan Phantom (see \cref{train_test_samp} for two representative samples). Given $N_\theta$ and $N_\tau$, we then simulate their corresponding sinograms $f$ based on standard discrete Radon transform \cite{austin2019simultaneous}.  Next we add $0.1\%$ Gaussian noise to each sinogram, respectively. This gives us our synthetic data, which we divide into $m = 20$ training samples and $m_{test} = 10$ testing samples.

We remark that in tomography, the \textit{number of projection angles}, $N_\theta$, is 
important, since it determines the amount of X-ray the sample is exposed to. We emphasize that the most 
challenging, yet common, cases in tomographic reconstruction are the ones with smaller $N_{\theta}$, 
due to the limits on X-ray exposure. We conduct numerical experiments for tomographic scans obtained for various $N_\theta$. For each choice, the selected number of angles are uniformly distributed in the range $[0,180]$. Note that, for each choice of $N_{\theta}$, a separate set of projection data is generated (for a batch of 30 samples), on which the learning and reconstructions are performed using our deep BONNet as discussed in \cref{train_phase_alg} and \cref{test_phase_alg}. 

We have undertaken two sets of experiments. In the first experiment, we fix $s = 0.4$ and learn $\lambda$. In the second experiment, we learn $\mu = (\lambda,s)$.

\subsubsection{Results of Experiment I: Learning $\lambda$, fixed $s = 0.4$.} \label{Exp_1}
\begin{figure}[htb]
\centering
\includegraphics[width = 0.49\textwidth, height = 0.6\textwidth]{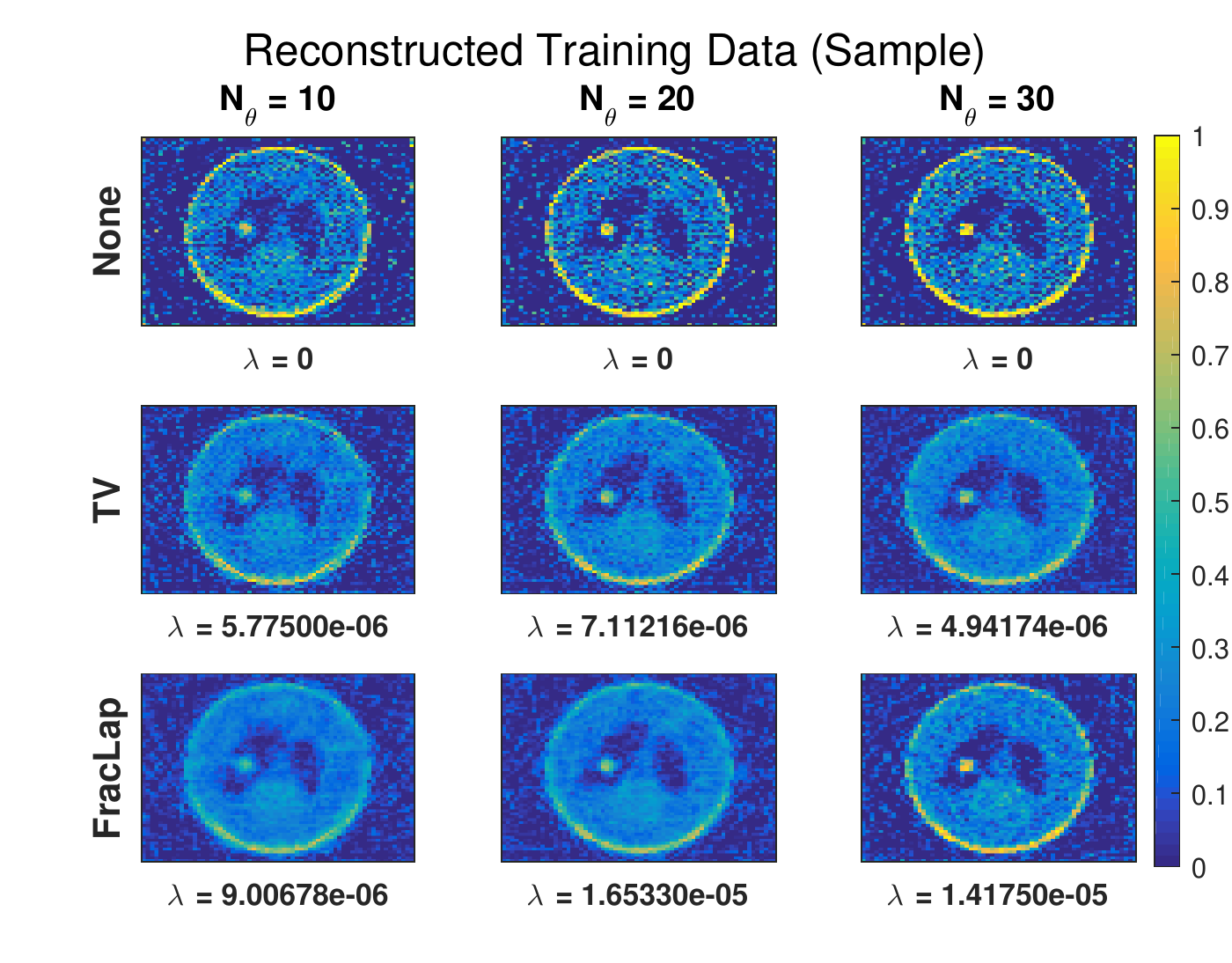} \includegraphics[width = 0.49\textwidth, height = 0.6\textwidth]{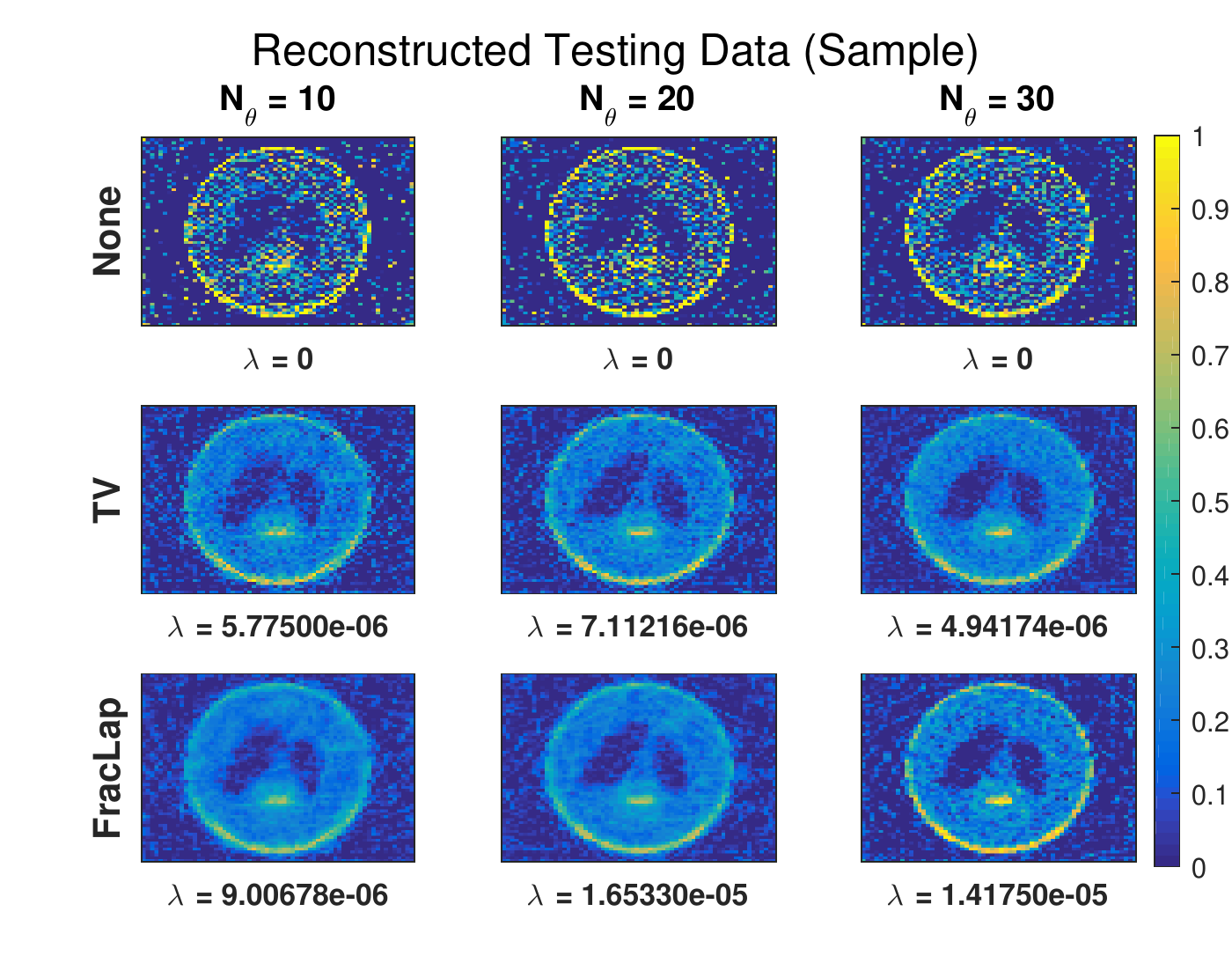}
\caption{Comparison of reconstructions based on various regularizers (\textit{rows}) and various number of tomographic projection angles (\textit{columns}) for data with $0.1\%$ Gaussian noise. The \textit{left} and \textit{right} panels corresponds to the solution at the last layer for two of the many distinct samples used during training and testing phases, respectively. The $\lambda$ values mentioned are the optimal values obtained from the deep BONNet training, which are then used for the reconstructions during the corresponding testing phase.}
\label{fig:FracTVCompnew}
\end{figure}

We now discuss the results of our experiments. In \cref{fig:FracTVCompnew}, we compare the reconstructions obtained from BONNet  with the true solution shown in \cref{train_test_samp}. 
The reconstructions are based on no regularization, total variation regularization, and the fractional Laplacian regularization for data with $0.1\%$ noise. The columns correspond to the number of projections angles used. We remark again that each choice of $N_{\theta}$ for a batch of training and testing data, corresponds to a distinct separate problem that we solve, as the dimensionality of ${\bf K}$ depends on $N_{\theta}$. The \textit{left panel} corresponds to the reconstruction of the \textit{training} data at the $nth$ iterate. Recall that at the training phase, $\{(u_{true}^{(i)},f_{train}^{(i)})\}_{i=1}^{m=20}$  are passed to the deep BONNet \cref{train_phase_alg}. The $\lambda$ values mentioned under each reconstruction are the corresponding optimal $\lambda_{none}^*,\lambda_{TV}^*,$ and $\lambda_{fracLap}^*$ that we learn during the training stage. Notice that $\lambda_{none}^*=0$ corresponds to no regularization. The \textit{right panel} corresponds to the reconstructions at the $n_{test}th$ layer of the testing phase. Recall that $\{(\lambda^*, f_{test}^{(i)})\}_{i=1}^{m_{test}=10}$ are passed to the deep BONNet at this stage of \cref{test_phase_alg}. 

From the reconstructions in \cref{fig:FracTVCompnew}, we observe that for the tomographic reconstruction problem, first of all, regularization is 
improving the quality of reconstructions. In the absence of regularization, the high intensity regions are preserved, but we lose information from regions of low intensity. On the other hand, TV and fractional Laplacian regularizations preserve the sample characteristics in the lower intensity regions of the sample. Fractional Laplacian gives reconstructions which are either better, or comparable to TV regularization. In addition, it does better at smoothing out the noise, and also in regaining comparatively more information in regions of low intensity, such as the dim circle on the lower side of the Phantom, e.g. for $N_{\theta} = 10$. This is especially important when we have limited data to reconstruct from.  
 We also recall that the Euler-Lagrange equation corresponding to the fractional Laplacian regularization is linear, and that of TV is non-linear. 

We also observe that for any given regularizer choice, the optimal $\lambda^*$ obtained for $N_{\theta}=10$ is similar to the one obtained for a larger $N_{\theta}$. Thus, to learn the regularization strength, even limited tomographic scan data suffices, and the same $\lambda^*$ could be used for reconstruction at the testing phase for any amount of available data, which can significantly save the offline training time. 

\begin{figure}[h!]
\centering
\includegraphics[width=0.49\textwidth, height = 0.4\textwidth]{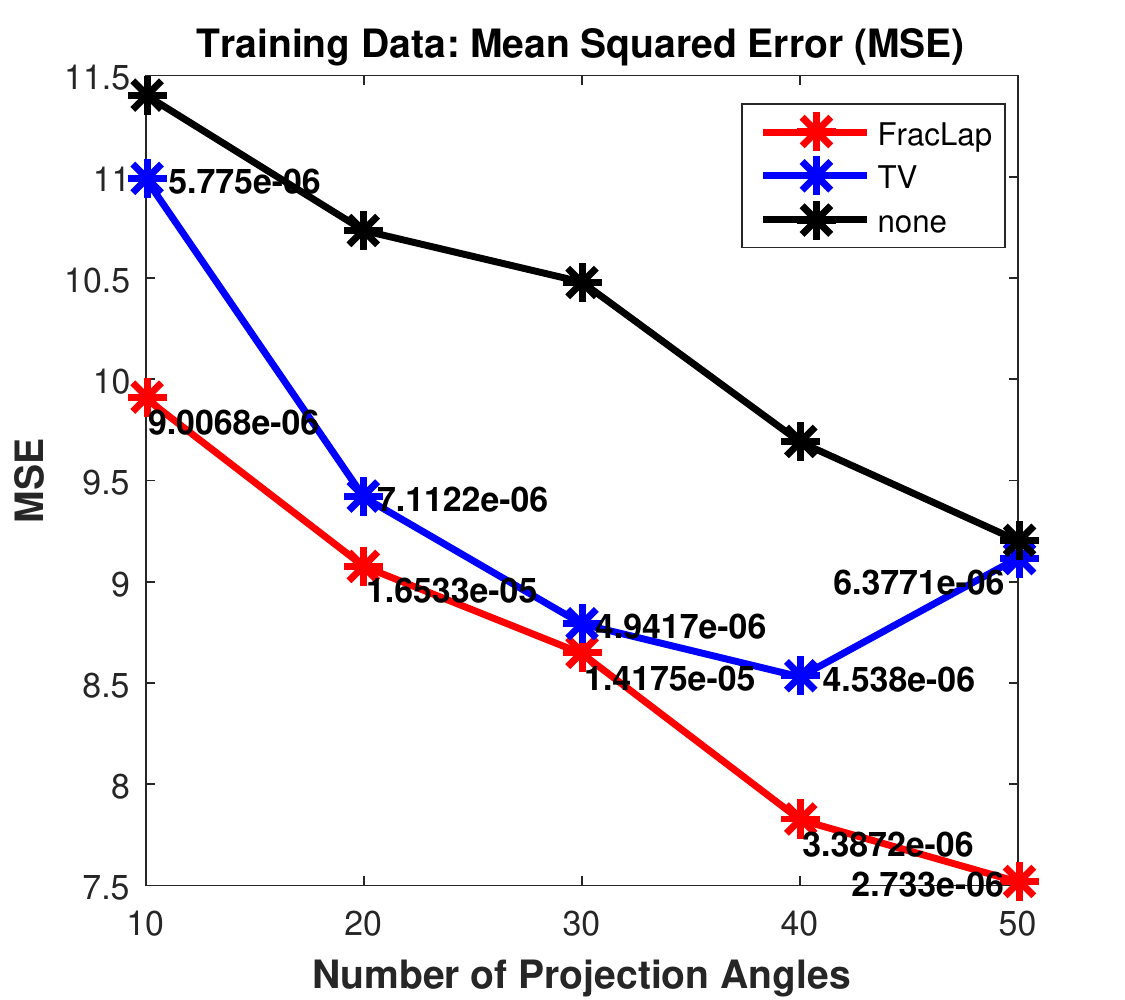} \includegraphics[width=0.49\textwidth, height = 0.4\textwidth]{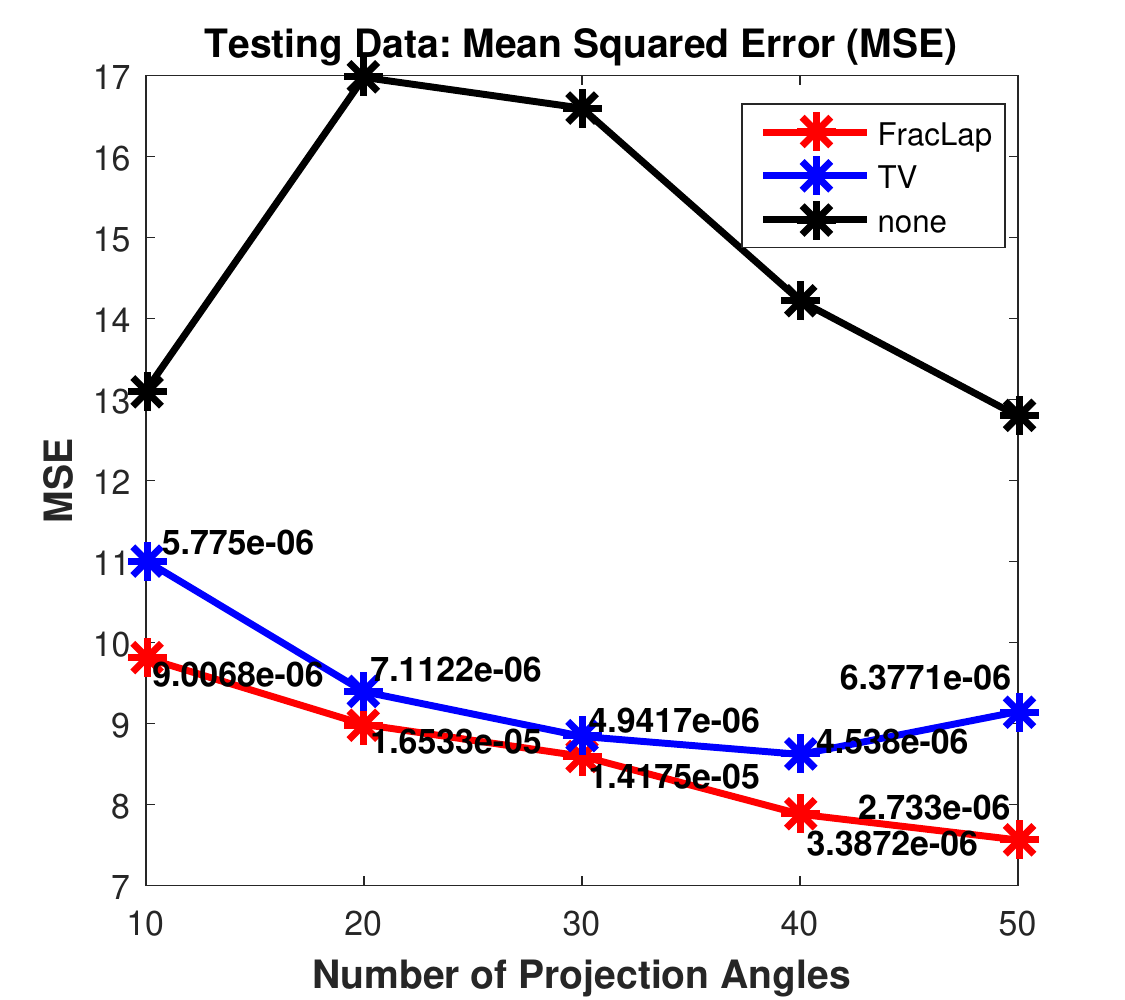}
\caption{We compare the mean-squared errors (MSE) for the solution, averaged over 20 training (respectively, 10 testing) samples (\textit{left}(respectively, right)), against various number of projection angles for the tomographic reconstruction problem. The solid \textit{black, blue} and \textit{red} lines corresponds to no regularization, total variation regularization, and fractional Laplacian regularization, respectively. For each experiment, the $\lambda^*$ learned from BONNet at the training phase is mentioned, which is in turn used for the reconstruction during training (\textit{left}) and testing (\textit{right}) phases. Smaller values of MSE correspond to better results, and fractional Laplacian outperforms the others. Note that $0.1\%$ Gaussian noise was added to the data $`f'$, and $s=0.4$ for fractional Laplacian.}
\label{fig:avg_norm_compar}
\end{figure}

\begin{figure}[h!]
\centering
\includegraphics[width=0.49\textwidth, height = 0.4\textwidth]{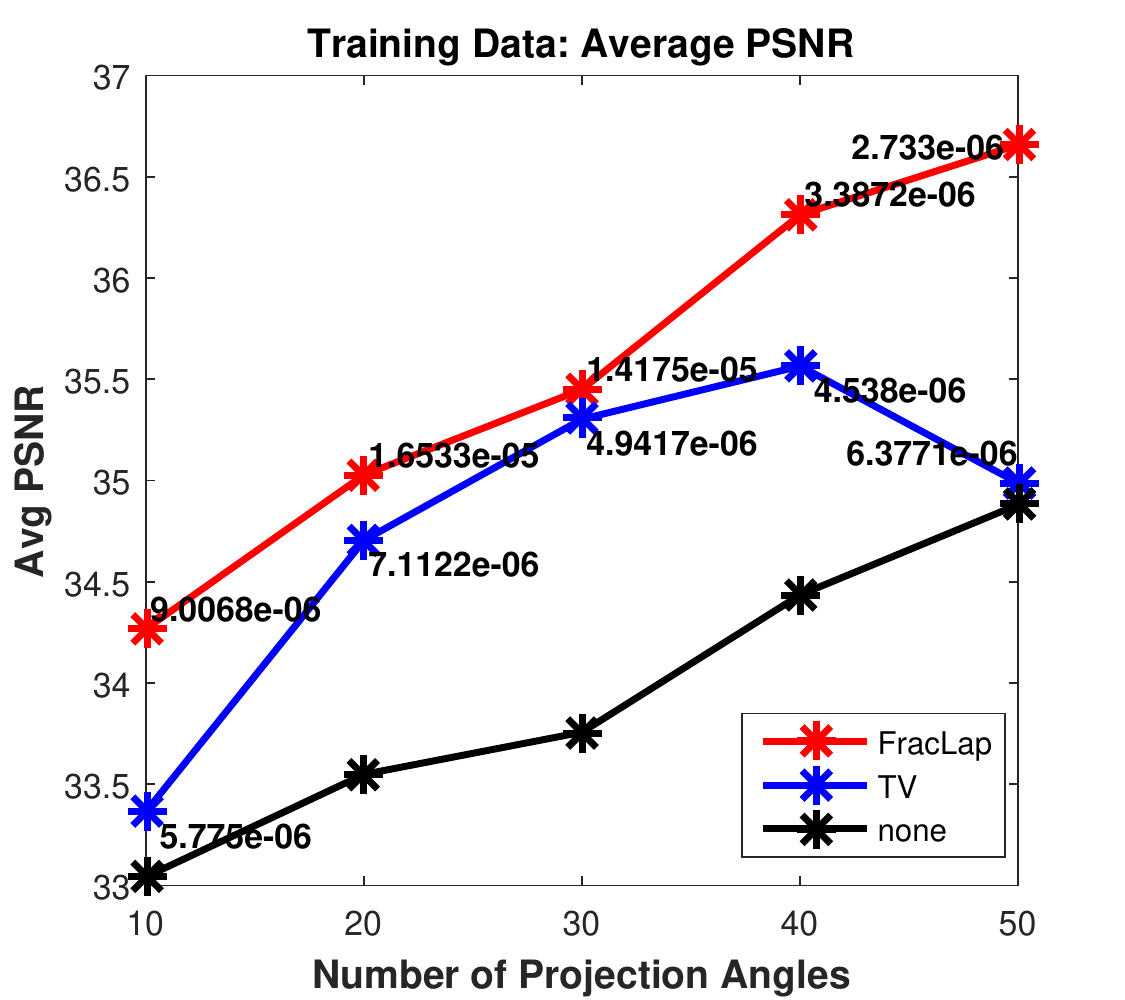}  \includegraphics[width=0.49\textwidth, height = 0.4\textwidth]{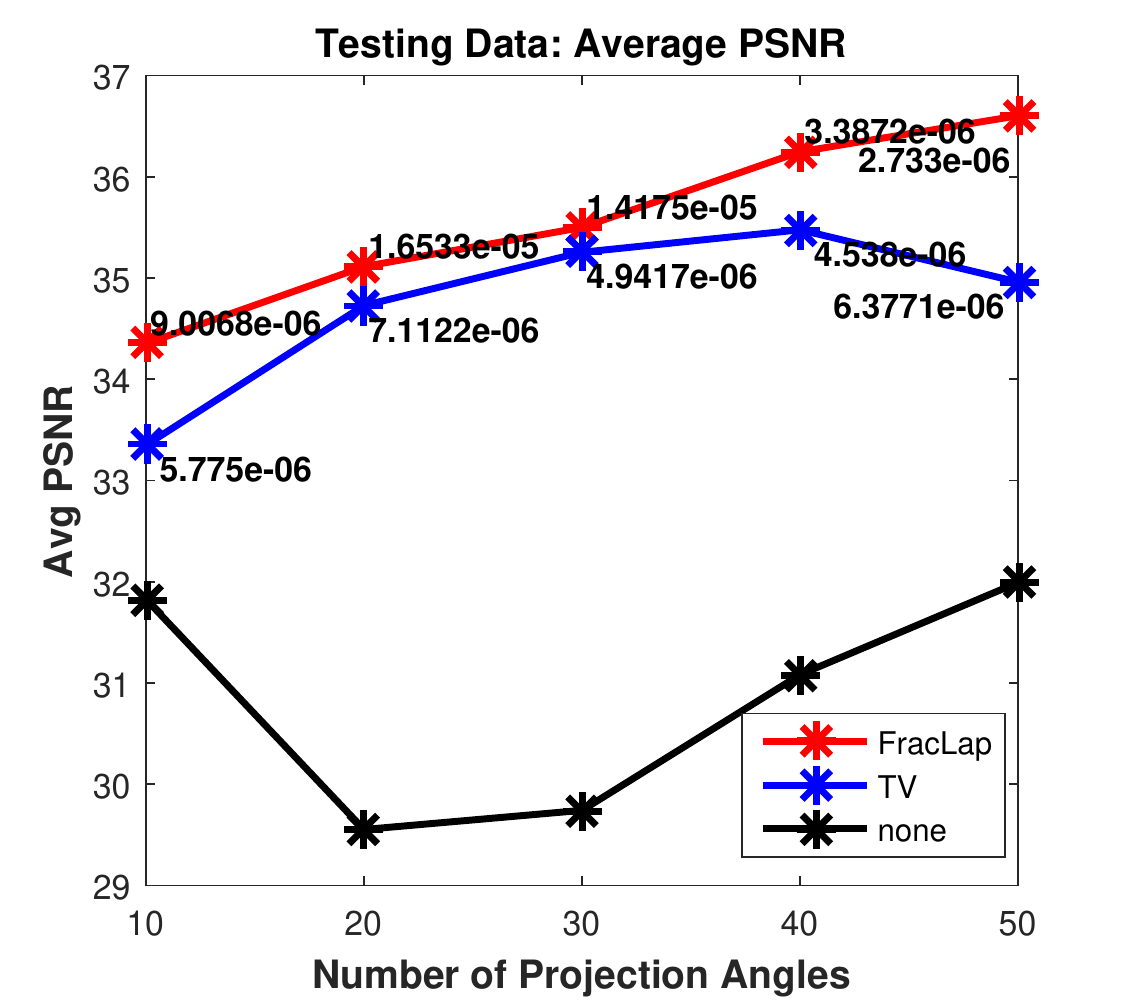}
\caption{We compare the peak signal-to-noise ratio (PSNR) for the solution, averaged over 20 training (respectively, 10 testing) samples (\textit{left}(respectively, right)), against various number of projection angles for the tomographic reconstruction problem. The solid \textit{black, blue} and \textit{red} lines corresponds to no regularization, total variation regularization, and fractional Laplacian regularization, respectively. For each experiment, the $\lambda^*$ learned from BONNet at the training phase is mentioned, which is in turn used for the reconstruction during training (\textit{left}) and testing (\textit{right}) phases. Larger values of PSNR correspond to better results, and fractional Laplacian outperforms the others. Note that $0.1\%$ Gaussian noise was added to the data $`f'$, and $s=0.4$ for fractional Laplacian.}
\label{fig:avg_psnr_compar}
\end{figure}

\begin{figure}[htb]
\centering
\includegraphics[width=0.49\textwidth, height = 0.4\textwidth]{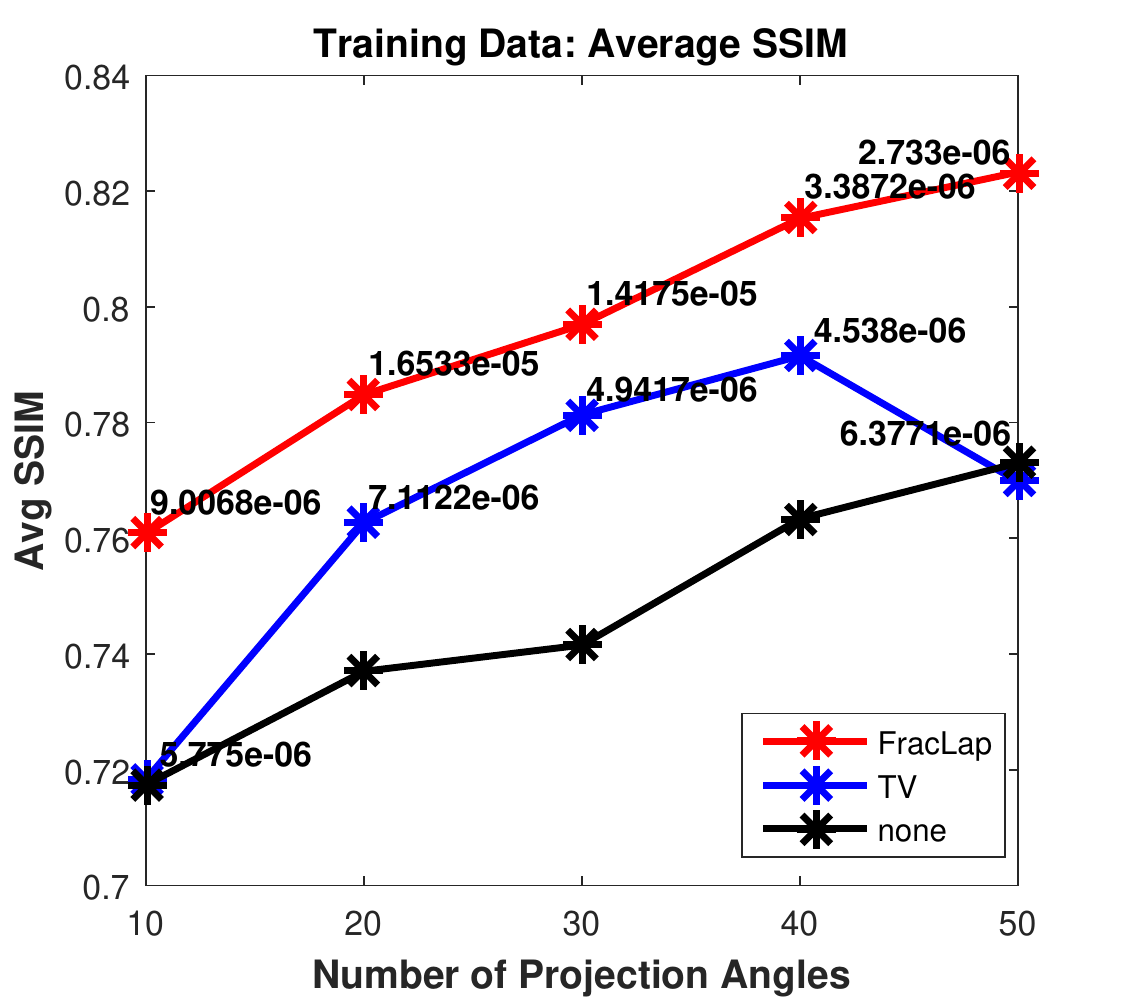}  \includegraphics[width=0.49\textwidth, height = 0.4\textwidth]{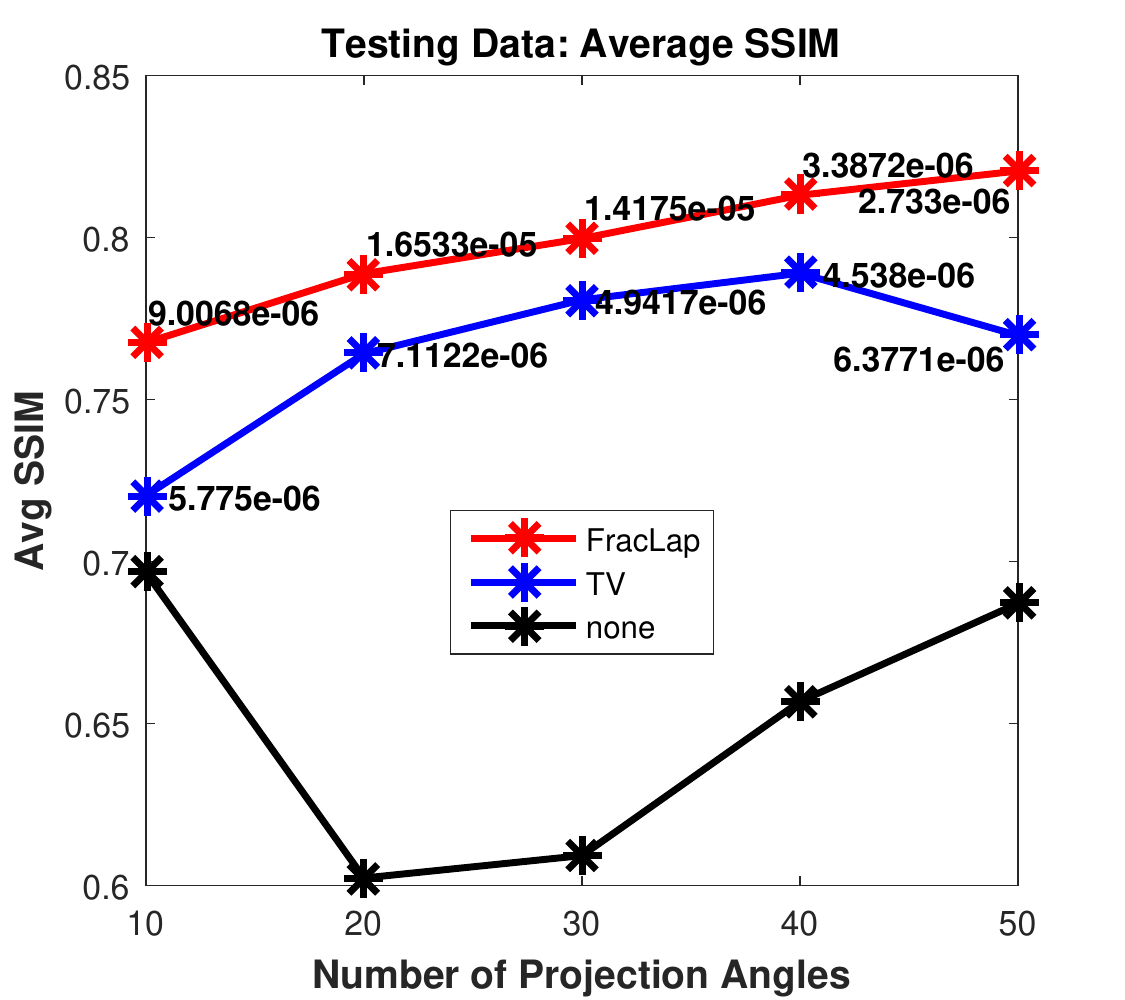}
\caption{We compare the peak structural similarity (SSIM) for the solution, averaged over 20 training (respectively, 10 testing) samples (\textit{left}(respectively, right)), against various number of projection angles for the tomographic reconstruction problem. The solid \textit{black, blue} and \textit{red} lines corresponds to no regularization, total variation regularization, and fractional Laplacian regularization, respectively. For each experiment, the $\lambda^*$ learned from BONNet at the training phase is mentioned, which is in turn used for the reconstruction during training (\textit{left}) and testing (\textit{right}) phases. Larger values of SSIM correspond to better results, and fractional Laplacian outperforms the others. Note that $0.1\%$ Gaussian noise was added to the data $`f'$, and $s=0.4$ for fractional Laplacian.}
\label{fig:avg_ssim_compar}
\end{figure}

For the experimental cases mentioned above, we measure the quality of reconstructions using metrics such as  the \textit{mean-squared error} (MSE) \cref{fig:avg_norm_compar}, \textit{Peak signal-to-noise ratio} (PSNR)  \cref{fig:avg_psnr_compar}, and \textit{structural similarity index} (SSIM) \cref{fig:avg_ssim_compar}, averaged over all the samples. For MSE, smaller values correspond to better results, and for PSNR and SSIM, larger values are better. Notice that for each metric, fractional Laplacian regularization outperforms the total variation regularization. 

We remark that the $\lambda$ values that we learn via deep BONNet are similar to those obtained by using a combination of the lowest error norm and L-curve; however, the parameter search via BONNet is automated. The  reconstructions obtained via Projected Gradient Descent are also similar to the ones obtained earlier \cref{fig:FracTVCompold} using the inexact truncated-Newton method for bound-constrained problem \cite{Nash200045}. We emphasize that one may use a different solver during the testing stage once $\lambda^*$ is obtained via BONNet training. 

\subsubsection{Results of Experiment II: Learning $\lambda$ and Fractional Exponent $`s$'. \label{numerics_s}}
We now train BONNet to learn both the fractional exponent $`s$' of the fractional Laplacian and the strength $\lambda$. We use the BONNet architecture using fractional Laplacian discussed in \cref{fracLap_archit}  
and use the same training and testing data as described in the previous example.
In \cref{lam_s_compar_test} we show comparisons of MSE, SSIM and PSNR for $N_{\theta} = \{ 10,20\}$ projection angles, respectively, for the reconstructions of the testing data. We compare the results with the fractional Laplacian case discussed in \cref{Exp_1}. In the case of $N_\theta = 10$, we obtain $(\lambda^*_{fracLap},s^*$) = (5.04417e-6, \ 0.5413) and in the case of $N_\theta = 20$, we obtain $(\lambda^*_{fracLap},s^*)$ = (8.53717e-6, \ 0.3799). The reconstructions of $u$  with $(\lambda^*_{fracLap},s^*)$ are visually comparable to the case of fractional Lapalcian in  \cref{fig:FracTVCompnew} and therefore they have been omitted. We observe that all the error metrics returned by BONNet are either comparable, or slightly better, than the ones obtained by BONNet for a fixed $`s$', discussed in \cref{Exp_1}. The advantage now is that we no longer need to tune the parameters manually. 
%

\begin{table}[htb] \label{lam_s_compar_test}
\begin{center}
\begin{scriptsize}
\begin{tabular}{|c|c|c|c|c|}
\hline
\textbf{Data}         & \multicolumn{4}{c|}{\textbf{Testing}}                                   \\ \hline
\textbf{$N_{\theta}$} & \multicolumn{2}{c|}{\textbf{$10$}} & \multicolumn{2}{c|}{\textbf{$20$}} \\ \hline
\textbf{Type}   & Experiment I   & Experiment II                 & Experiment I   & Experiment II  \\ \hline
\textbf{$(\lambda,s)$}    & $(9.00678e-6,0.4)$  & $(5.04417e-6,0.5413)$       & $(1.65330e-5,0.4)$  & $(8.53717e-6,0.3799)$  \\ \hline
\textbf{MSE}          & $9.8099$      & $9.7743$           & $8.9872$      & $8.6961$           \\ \hline
\textbf{SSIM}         & $0.7675$      & $0.7738$           & $0.7888$      & $ 0.7950$          \\ \hline
\textbf{PSNR}         & $34.3513$     & $34.3831$          & $35.1123$     & $35.3973$          \\ \hline
\end{tabular}
\end{scriptsize}
\caption{Comparison of average MSE, SSIM and PSNR for tomographic reconstructions obtained via BONNet using the fractional Laplacian regularization for two projection angles. In Experiment I, we fix $s=0.4$ and learn $\lambda^*$ via BONNet, and in Experiment II we learn the $(\lambda^*,s^*)$ pair. The results shown are for the testing dataset. Notice that the search for $\mu^* = (\lambda^*,s^*)$ in Experiment II is now fully automated and the results are better or comparable to the previous case.
}
\end{center}
\end{table}

\section{Discussion.}

In this work, we consider a  general regularized regression model for inverse problems. This model can incorporate the underlying physics (defined by the operator $K$), in addition to the prior knowledge of the solution in the regularization term. However, to fully explore the potential of this generalized model, an optimal choice of the type of regularizer, as well as the  regularization strength, is inevitable.

We have used fractional Laplacian as a regularizer on tomographic reconstruction problems. Previously, this has been used in image denoising. 
The key benefit of using this regularization is that the corresponding Euler-Lagrange equations are \textit{linear}, as opposed to the \textit{nonlinear} and possibly \textit{degenerate} Euler-Lagrange equations for the popular total variation regularization. 

To address the challenge of finding the optimal regularization strength, we introduce a dedicated deep BONNet architecture to learn the regularization parameters for any choice of regularizer. 
We show an analogy of the regularization function to the activation function in a standard neural network, which provides a theoretical guidance in terms of choosing an optimal activation function.
In addition to the regularization strength $\lambda$, BONNet can also learn the exponent $`s$' for the fractional Laplacian regularization. 

Next, we demonstrate the benefit 
of our proposed deep BONNet on the tomographic reconstruction problem. 
We first conduct experiments to learn only $\lambda$ with a fixed $`s$'. We have observed that fractional Laplacian regularization gives comparable or better reconstructions compared to the total variation regularization. Especially for the noisy and limited data 
($N_{\theta} = 10$), fractional Laplacian regularization outperforms the total variation regularization. 
In contrast to the standard machine learning architectures with fixed number of layers, 
our network favors a variable number of layers (depth) which is dictated by the convergence to the solution of the optimization problem. Thus, the number of layers in the network can be different for different samples and different regularizers. We also demonstrate the capability of our proposed BONNet in terms of learning 
the optimum $(\lambda^*_{fracLap},s^*)$ pair for the fractional Laplacian regularizer, and this indicates the flexibility of our proposed network to learn non-standard parameters.

\bibliographystyle{siamplain}
\bibliography{refs}
\end{document}